\documentclass[leqno]{lmcs}
\pdfoutput=1

\usepackage{lastpage}
\lmcsdoi{17}{1}{17}
\lmcsheading{}{\pageref{LastPage}}{}{}%
{Jan.~15,~2020}{Feb.~18,~2021}{}

\usepackage[utf8]{inputenc}


\usepackage{stmaryrd,amssymb,amsmath,amsthm,url}

\usepackage{environ}

\makeatletter
\NewEnviron{Lalign}{\tagsleft@true\begin{align}\BODY\end{align}}
\makeatother

\usepackage{environ}
\makeatletter
\NewEnviron{Ralign}{\tagsleft@false\begin{align}\BODY\end{align}}
\makeatother

\newcommand{\sumi}{\textstyle{\sum}}
\newcommand{\apb}{\app b}
\newcommand{\apbz}{\apb 0}
\newcommand{\apbo}{\apbu}
\newcommand{\apbu}{\apb 1}
\newcommand{\apd}{\app d}
\newcommand{\apt}{\app t}

\newcommand{\conca}[1]{\hspace{1mm}\hat{\raisebox{-,4ex}{\footnotesize{\it #1}}}\hspace{1mm}}
\newcommand{\concu}{\conca u}
\newcommand{\concb}{\conca b}
\newcommand{\concd}{\conca d}
\newcommand{\conct}{\concd}

\newcommand{\Nat}{{\mathbb N}}
\newcommand{\NF}{\mathit{N}}
\newcommand{\Int}{{\mathbb Z}}
\newcommand{\PNF}{{\NF^+}}
\newcommand{\NNF}{{\NF^-}}

\newcommand{\dname}[1]{D_{#1}}
\newcommand{\nname}[1]{\mathit{Nat}_{#1}} 
\newcommand{\zname}[1]{\mathit{Int}_{#1}} 

\newcommand{\app}[1]{\hspace{.4mm}{:_{\hspace{.2mm}#1}}\hspace{.4mm}}
\newcommand{\apu}{\app u}
\newcommand{\apue}{\apu1}

\begin{document}

\title{Datatype defining rewrite systems for naturals and integers} 

\author[J.A. Bergstra]{Jan A.\ Bergstra}

\author[A. Ponse]{Alban Ponse}
\address{Informatics Institute, section Theory of Computer Science, University of Amsterdam}
\email{j.a.bergstra@uva.nl}
\email{a.ponse@uva.nl}

\begin{abstract}
A datatype defining rewrite system (DDRS) is an algebraic (equational) specification 
intended to specify a datatype. When interpreting the equations from left-to-right, 
a DDRS defines a term rewriting system that must be ground-complete.
First we define two DDRSs for the ring of 
integers, each comprising twelve rewrite rules, and prove their ground-completeness.
Then we introduce natural number and integer arithmetic specified according to unary view, 
that is, arithmetic based on a postfix unary append constructor (a form of tallying).
Next we specify arithmetic based on two other views: binary and decimal notation. 
The binary and decimal view have as their characteristic that each normal form resembles 
common number notation, that is, either a digit, or a string of digits without leading zero, 
or the negated versions of the latter.
Integer arithmetic in binary and decimal notation is based on (postfix) digit append functions. 
For each view we define a DDRS, and in each case the resulting datatype 
is a canonical term algebra that extends a corresponding canonical term algebra for natural numbers.
Then, for each view, we consider an alternative DDRS based on tree constructors that yields comparable 
normal forms, which for that view admits expressions that are algorithmically more involved. 
For all DDRSs considered, ground-completeness is proven.
\end{abstract}

\maketitle

\section{Introduction}
We specify natural number arithmetic and integer arithmetic by algebraic specifications,
according to three different ``views'': 
unary, binary, and decimal notation. 
This paper is based on the specifications for natural numbers from~\cite{Bergstra2014n} that 
define addition and multiplication, and we follow the same strategy to develop these different views. 
Each of the specifications provided is a so-called
DDRS (datatype defining rewrite system) and consists of a number of equations that define
a {term rewriting system} (TRS) when interpreting the equations from left-to-right. 
A DDRS must be ground-complete, that is, terminating and ground-confluent; 
for some general information on TRSs see e.g.~\textsc{Terese}~\cite{Terese}.

The \emph{unary view} in~\cite{Bergstra2014n} is defined by a DDRS for which 0 and 
successor terms are the normal forms. The unary view
is also used to provide a semantic specification of binary and decimal notation, 
using operator symbols for appending a digit. 
These two positional notations were modified
with respect to conventional notations in such a way that syntactic confusion between these 
notations cannot arise. 
It seems to be the case that for the unary view the specification of the integers is entirely
adequate, whereas all subsequent specifications for binary and decimal view may provide no more
than a formalisation of a topic which must be somehow understood before taking notice of that 
same formalisation. It remains to be seen to what extent the DDRSs for the unary case may serve 
exactly that expository purpose.
Furthermore, each of these DDRSs contains equations for rewriting 
constructor terms in one of the other views to a term in the DDRS's
view, e.g., the DDRSs for the unary view contain the 
equation $1=S(0)$ for the constant 1 in binary view and in decimal view.
The definitions of these DDRSs are geared towards obtaining comprehensible specifications of 
natural number and integer arithmetic in binary and decimal view. The successor 
function $S(x)$ and predecessor function $P(x)$
appeared to be instrumental auxiliary 
functions for this purpose, thereby justifying the incorporation of the unary view
as a separate view.

This paper constitutes a further stage in the development of a family of arithmetical 
datatypes with corresponding specifications. 
The resulting DDRSs incorporate different views on the same abstract datatype
(ADT), where an ADT may be understood as the isomorphism class of its instantiations.
The datatypes considered in~\cite{Bergstra2014n} are so-called 
canonical term algebras (we further discuss these in Section~\ref{sec:2}).

The strategy of this work is somewhat complicated: on the one hand we 
look for specifications that may genuinely be considered introductory, 
that is, descriptions that can be used to construct the datatype at hand 
for the first time in the mind of a person.
On the other hand awareness of the datatype in focus may be needed to 
produce an assessment of the degree of success achieved in the direction of the first objective.
The described models aim to represent the natural and integer numbers. 
The question whether this is really 
the case depends on one's conception of the natural numbers as well as on the requirements one maintains of 
the notion of proof. As an independent foundational question, this question is neither posed nor answered 
in this paper, though the arguments in favour of the recognition of representations of natural and 
integer numbers in the given specifications are in line with conventional work on 
abstract datatypes. For such arguments ground-completeness is clearly an important aspect, 
and is the key focus of the current paper.
The most closely related work seems to be that of \textsc{Walters \& Zantema}~\cite{WZ95};
in Section~\ref{sec:6} we briefly discuss a comparison with this work.

\begin{table}
\hrule
\begin{minipage}[t]{0.48\linewidth}\centering
\begin{Lalign}
\tag*{[CR1]}
\hspace{-22mm}
x + (y + z) &= (x+y) + z
\hspace{-22mm}
\\
\tag*{[CR2]}
x+y &= y + x\\
\tag*{[CR3]}
x+0 &= x\\
\label{CR4}
\tag*{[CR4]}
\hspace{-22mm}
x + (-x) &= 0
\end{Lalign}
\end{minipage}
\hspace{-8mm}
\begin{minipage}[t]{0.556\linewidth}\centering
\begin{Lalign}
\tag*{[CR5]}
(x \cdot y) \cdot z &= x \cdot (y \cdot z) \\
\tag*{[CR6]}
x \cdot y &= y \cdot x\\
\tag*{[CR7]}
1\cdot x&=x\\
\tag*{[CR8]}
x \cdot (y + z) & = (x \cdot y) + (x \cdot z)
\end{Lalign}
\end{minipage}\vspace{2mm}
\hrule
\caption{Axioms for commutative rings}
\label{tab:CR}
\end{table}

The paper is structured as follows. In Section~\ref{sec:2}
we start with DDRSs for commutative rings over the signature 
$\Sigma_r=\{0,1,-(\_),+,\cdot\}$, 
which are defined by the axioms Table~\ref{tab:CR} (we shall often write $-t$ for $-(t)$). These axioms
characterise integer arithmetic, while leaving open how 
numbers are represented (apart from the constants $0,1\in\Sigma_r$).
For the ADT $\Int$ determined by these axioms we introduce two simple DDRSs
that both have the same normal forms.
In Section~\ref{sec:3} we define two DDRSs for the unary view that 
relate to the DDRSs for the ring of integers, with numbers represented in unary
view, that is, in a numeral system based on the constant 0 and a \emph{unary append} 
constructor (a form of tallying).
Moreover, a part of each of these DDRSs specifies natural number arithmetic.
In Section~\ref{sec:4} we define DDRSs for natural number and integer arithmetic
that follow the approach in~\cite{Bergstra2014n}.
These DDRSs combine unary view with binary and 
decimal view and are based on \emph{digit append} constructors.
In Section~\ref{sec:5} we define similar DDRSs that employ
\emph{digit tree} constructors instead, which are algorithmically more 
involved.
In Section~\ref{sec:6} we come up with some conclusions. 
The paper contains two appendices with detailed proofs.

This paper subsumes, and thereby also replaces and improves on, 
our earlier paper \cite{BP16a}. The main differences are that the DDRSs defined in 
Tables~\ref{tab:r1} -- \ref{tab:int2} are new, and that the DDRSs defined
in all remaining tables are simplified and contain fewer rules, except  
those in Table~\ref{tab:intbview}. 
All termination proofs were found by the tool AProVE~\cite{AProVE},
downloads of these proofs are available at \url{https://arxiv.org/src/1608.06212/anc/}.

\medskip

\textit{This paper has been written for the occasion of the retirement of Jos Baeten, lately as 
the director of the CWI in Amsterdam. For Alban, Jos was his highly valued
second promotor at the University of Amsterdam. For Jan, in addition to being a former UvA 
colleague, Jos has been a highly respected coauthor of many papers. Jos has been participating in 
the ACP process algebra project from a quite early stage and the outcome of our joint work has 
been important for Jan's work ever since.}

\section{DDRSs for the ring of integers}
\label{sec:2}
In this section we introduce two DDRSs that both specify the ring of integers.
In Section~\ref{sec:2.1} we first give a general definition of a DDRS and discuss in what way
it specifies a datatype, and then provide a DDRS for the ring of integers.
In Section~\ref{sec:2.2} we consider an alternative DDRS for the ring of integers that
is deterministic with respect to rewriting a sum of two nonnegative closed normal forms.

\subsection{A DDRS for the ring of integers}
\label{sec:2.1}

We start with a formal definition of a DDRS.
\begin{defi}
Given a many-sorted signature $\Sigma$ and a finite set $E$ of equations over $\Sigma$, 
a specification $(\Sigma, E)$ is
a \textbf{DDRS} (Dataype Defining Rewrite System) if for each sort $S$ in $\Sigma$
the following two requirements are satisfied: 
\begin{enumerate}
\item there is a closed term
of sort $S$, thus, $S$ is inhabited,
\item the
equations in $E$ over sort $S$ when interpreted from left to right
define a ground-complete TRS.
\end{enumerate}
\end{defi}

The datatypes specified by the forthcoming DDRSs are \emph{canonical term algebras},
which means that carriers are non-empty sets of closed terms, and
for each congruence class of closed terms, a unique 
representing term is chosen, and this set of representing closed terms, the \emph{normal forms},
is closed under taking subterms.

\medskip

In this paper we will consider single-sorted DDRSs for integer arithmetic.
We require that for each such DDRS, $0$ is a constant in its signature
and for each closed term $t$ both
$t+0=t$ and $t+(-t)=0$ hold (so, e.g.\ exchanging the
roles of $0$ and $1$ in such a DDRS is not permitted). 

\medskip

In~\cite{BP16,BP16a} we defined a DDRS consisting of fifteen equations
for the ADT $\Int$ (which is
determined by the axioms for commutative rings in Table~\ref{tab:CR}). 
Normal forms for this DDRS are 0 for zero,
the positive normal forms~1 and $t+1$ with $t$ a positive normal form,
and the negations of positive normal forms, thus $-t$ for each positive normal form $t$.
Clearly, two different closed normal forms have distinct values in $\Int$.
The DDRS $\dname 1$ in Table~\ref{tab:r1} over the signature $\Sigma_r$ defines the datatype $\Int_r$
that has the same normal forms and is hence the same CTA. 

\begin{table}
\centering
\hrule
\begin{minipage}[t]{0.49\linewidth}\centering
\begin{Lalign}
\label{R1}
\tag*{\textup{[R1]}}
x+0 &=x
\\
\label{R2}
\tag*{\textup{[R2]}}
0+x &=x
\\
\label{R3}
\tag*{\textup{[R3]}}
x + (y+z) &= (x+y)+z
\\[4mm]
\label{R4}
\tag*{\textup{[R4]}}
x \cdot 0 &=0\\
\label{R5}
\tag*{\textup{[R5]}}
x \cdot 1 &=x\\
\label{R6}
\tag*{\textup{[R6]}}
x \cdot (y+z) &=(x\cdot y)+(x\cdot z)
\end{Lalign}
\end{minipage}
\hspace{-9mm}
\begin{minipage}[t]{0.55\linewidth}\centering
\begin{Lalign}
\label{R7}
\tag*{\textup{[R7]}}
-0&=0
\\
\label{R8}
\tag*{\textup{[R8]}}
(-1)+1&=0\\
\label{R9}
\tag*{\textup{[R9]}}
(-(x+1))+1&=-x
\\
\label{R10}
\tag*{\textup{[R10]}}
-(-x) &=x
\\[2mm]
\label{R11}
\tag*{\textup{[R11]}}
x+(-y)&=-((-x)+y)
\\[2mm]
\label{R12}
\tag*{\textup{[R12]}}
x \cdot (-y) &=-(x\cdot y)
\end{Lalign}
\end{minipage}\vspace{4mm}
\hrule
\caption{The DDRS $\dname1$ for the datatype~$\Int_r$ that specifies
the ring of integers}
\label{tab:r1}
\end{table}

Clearly, all equations of $\dname 1$ are semantic consequences
of the axioms for commutative rings (Table~\ref{tab:CR}).
The difference between $\dname1$ in Table~\ref{tab:r1} and the DDRS 
for $\Int_r$ defined in~\cite{BP16,BP16a} is that equation~\ref{R11} replaces the four 
equations
\begin{Lalign}
\tag*{[r5]}
1+(-1)&=0,\\
\tag*{[r6]}
(x+1)+(-1)&=x,
\\
\tag*{[r7]}
x+(-(y+1))&=(x+(-y))+(-1),
\\
\tag*{[r11]}
(-x)+(-y)&=-(x+y).\hspace{8cm}
\end{Lalign}

\begin{thm}
\label{thm:2.1}
The DDRS $\dname1$ for $\Int_r$ defined in Table~\ref{tab:r1} is ground-complete.
\end{thm}

\begin{proof}
The AProVE tool~\cite{AProVE} finds that this DDRS is terminating.
For ground-conflu\-ence, see Corollary~\ref{cor:2.2.2}.
\end{proof}

\subsection{An alternative DDRS for the ring of integers}
\label{sec:2.2}
The DDRS $\dname1$ can be simplified by instantiating some of its equations.
In Table~\ref{tab:r2} we provide the 
DDRS $\dname2$ that also specifies the datatype $\Int_r$, where the
differences with $\dname1$ show up in the tags: equations~\ref{Rm2}, \ref{Rm3}
and~\ref{Rm6} replace \ref{R2}, \ref{R3} and~\ref{R6}, respectively. 

\begin{table}
\centering
\hrule
\begin{minipage}[t]{0.49\linewidth}\centering
\begin{Lalign}
\tag*{\ref{R1}}
x+0 &=x
\\
\label{Rm2}
\tag*{\textup{[R2$'$]}}
0+1 &=1
\\
\label{Rm3}
\tag*{\textup{[R3$'$]}}
x + (y+1) &= (x+y)+1
\\[4mm]
\tag*{\ref{R4}}
x \cdot 0 &=0\\
\tag*{\ref{R5}}
x \cdot 1 &=x\\
\label{Rm6}
\tag*{\textup{[R6$'$]}}
x \cdot (y+1) &=(x\cdot y)+x
\end{Lalign}
\end{minipage}
\hspace{-4mm}
\begin{minipage}[t]{0.52\linewidth}\centering
\begin{Lalign}
\tag*{\ref{R7}}
-0&=0
\\
\tag*{\ref{R8}}
(-1)+1&=0\\
\tag*{\ref{R9}}
~~~(-(x+1))+1&=-x
\\
\tag*{\ref{R10}}
-(-x) &=x
\\[2mm]
\tag*{\ref{R11}}
x+(-y)&=-((-x)+y)
\\[2mm]
\tag*{\ref{R12}}
x \cdot (-y) &=-(x\cdot y)
\end{Lalign}
\end{minipage}\vspace{4mm}
\hrule
\caption{The DDRS $\dname2$ for the datatype~$\Int_r$ that specifies
the ring of integers}
\label{tab:r2}
\end{table}

\begin{thm}
\label{thm:2.2}
The DDRS $\dname2$ for $\Int_r$ defined in Table~\ref{tab:r2} is ground-complete.
\end{thm}

\begin{proof}
The AProVE tool~\cite{AProVE} finds that this DDRS is terminating, so it
remains to be proven that $\dname2$ is ground-confluent.
Define the set $\NF$ of closed terms over $\Sigma_r$ as follows: 
\begin{align*}
\NF&=\{0\}\cup\PNF\cup\NNF,\\
\PNF&= \{1\}\cup\{t+1\mid t\in\PNF\},\\
\NNF&=\{-t\mid t\in\PNF\}.
\end{align*}
It immediately follows that if $t\in\NF$, then $t$ is a normal form (no rewrite step
applies). 
In order to prove ground-confluence of the DDRS $\dname2$ it suffices to show that 
for each closed term $t$ over $\Sigma_r$, either $t\in\NF$ or $t$ 
has a rewrite step, so that each normal form is in $\NF$. 
This is sufficient because $\dname2$ is terminating, 
all equations of $\dname 2$ are semantic consequences
of the axioms for commutative rings, and distinct closed 
normal forms have distinct values in $\Int$.

We prove this by structural induction on $t$. 
The base cases $t\in\{0,1\}$ are trivial. 
For the induction step we have to consider three cases:
\begin{enumerate}
\item
Case $t=-r$. Assume that $r\in\NF$ and apply case distinction on $r$:
\begin{itemize}
\item
if $r=0$, then $t$ has a rewrite step by equation~\ref{R7},
\item
if $r\in\PNF$, then $t\in\NF$, 
\item 
if $r\in\NNF$, then $t$ has a rewrite step by equation~\ref{R10}.
\end{itemize}
\item
Case $t=u+r$. Assume that $u,r\in\NF$ and apply case distinction on $r$:
\begin{itemize}
\item
if $r=0$, then $t$ has a rewrite step by equation~\ref{R1},
\item
if $r=1$, then apply case distinction on $u$: 
\begin{itemize}
\item
if $u=0$, then $t$ has a rewrite step by equation~\ref{Rm2},
\item
if $u\in\PNF$, then $t\in\NF$,
\item 
if $u=-1$, then $t$ has a rewrite step by equation~\ref{R8},
\item
if $u=-(u'+1)$, then $t$ has a rewrite step by equation~\ref{R9},
\end{itemize}
\item 
if $r=r'+1$, then $t$ has a rewrite step by equation~\ref{Rm3},
\item 
if $r=-r'$ with $r'\in\PNF$, then $t$ has a rewrite step by equation~\ref{R11}.
\end{itemize}

\item
Case $t=u\cdot r$. Assume that $u,r\in\NF$ and apply case distinction
on $r$:
\begin{itemize}
\item
if $r=0$, then $t$ has a rewrite step by equation~\ref{R4}, 
\item
if $r=1$, then $t$ has a rewrite step by equation~\ref{R5}, 
\item 
if $r=r'+1$, then $t$ has a rewrite step by equation~\ref{Rm6}, 
\item
if $r=-r'$ with $r'\in\PNF$, then $t$ has a rewrite step by equation~\ref{R12}.
\qedhere
\end{itemize}
\end{enumerate}
\end{proof}

\begin{cor}
\label{cor:2.2.2}
The DDRS $\dname1$ for $\Int_r$ defined in Table~\ref{tab:r1} is ground-confluent.
\end{cor}

\begin{proof}
It suffices to consider the 
proof of Theorem~\ref{thm:2.2} and to observe that each rewrite step by one
of the equations~\ref{Rm2}, \ref{Rm3}, and~\ref{Rm6} implies a rewrite step of
the associated equation in $\dname1$, which implies ground-confuence. 
\end{proof}

A particular property of $\dname2$ concerns the addition of two nonnegative normal forms. 

\begin{prop}
\label{prop:2.4}
With respect to addition of two nonnegative closed normal forms, 
the DDRS $\dname2$ in Table~\ref{tab:r2} is \textbf{deterministic}, 
that is, for nonnegative closed normal forms $t,t'$, in each state of rewriting of $t+t'$ 
to its normal form, only one equation (rewrite rule) applies.
\end{prop}

\begin{proof}
By structural induction on $t'$. For $t'\in\{0,1\}$ this is immediately clear.

If $t'=r+1$, then the only possible rewrite step is 
$t+(r+1)\stackrel{\text{\ref{Rm3}}}\longrightarrow(t+r)+1$. 
If $r=1$, this is a normal form, and if $r=r'+1$, the only redex in 
$(t+(r'+1))+1$ is in $t+(r'+1)$, and by induction the latter rewrites deterministically
to some normal form $u$. 

Consider this reduction:
$t+(r'+1)\to u'\to\!\!\!\!\to u$ 
(where $\to\!\!\!\!\to$ denotes zero or more rewrite steps). 
Then $(t+(r'+1))+1\to u'+1\to\!\!\!\!\to u+1$, and in each state of
this reduction the rightmost addition 
$...+1$ does not establish a new redex according to equations~\ref{R1}, \ref{Rm2}
and~\ref{Rm3}, and 
results in $u+1$, which also is a normal form. 
\end{proof}

It is clear that for example $1+t'$ with $t'$ a closed negative normal form also rewrites 
deterministically in $\dname2$, and that
this determinism is not preserved for $t+t'$ if $t=0$, e.g.,

{\small
\begin{align*}
&0+(-(r+1))\stackrel{\text{\ref{R11}}}\longrightarrow
-((-0)+(r+1))\left[\begin{array}{l}
\stackrel{\text{\ref{Rm3}}}\longrightarrow-(((-0)+r)+1)
\stackrel{\text{\ref{R7}}}\longrightarrow\\[2mm] 
\stackrel{\text{\ref{R7}}}\longrightarrow~ {-}(0+(r+1))
\hfill\stackrel{\text{\ref{Rm3}}}\longrightarrow\end{array}\right]
{-}((0+r)+1),
\end{align*}
}
or if $t$ is a closed negative normal form $(-t')$, e.g.,

{\small
\begin{align*}
&(-t')+(-(r+1))\stackrel{\text{\ref{R11}}}\longrightarrow
-((-(-t'))+(r+1))\left[\begin{array}{l}
\stackrel{\text{\ref{Rm3}}}\longrightarrow-(((-(-t'))+r)+1)
\stackrel{\text{\ref{R10}}}\longrightarrow\\[2mm] 
\stackrel{\text{\ref{R10}}}\longrightarrow~ {-}(t'+(r+1))
\hfill\stackrel{\text{\ref{Rm3}}}\longrightarrow\end{array}\right]
{-}((t'+r)+1).
\end{align*}
}

However, our interest in deterministic reductions concerns nonnegative
closed normal forms and we return to this point in the next section.
Finally, note that with respect to multiplication of two nonnegative normal forms, 
the DDRS $\dname 2$ is not deterministic:

{\small
\[
0\cdot(1+1)\stackrel{\text{\ref{Rm6}}}\longrightarrow
(0\cdot1)+0\left[\begin{array}{l}
\stackrel{\text{\ref{R4}}}\longrightarrow0+0
\stackrel{\text{\ref{R1}}}\longrightarrow\\[2mm] 
\stackrel{\text{\ref{R1}}}\longrightarrow ~0\cdot1
\hfill\stackrel{\text{\ref{R4}}}\longrightarrow\end{array}\right]
0.
\]
}

\section{DDRSs for natural number and integer arithmetic in unary view}
\label{sec:3}
Given the signature 
$\{0,1,+,\cdot\}$, natural number arithmetic can be characterised 
by the axioms in Table~\ref{tab:Nat} and we write $\Nat$ for the ADT captured by these axioms.
In the remainder of this paper we will also discuss  
single-sorted DDRSs for natural number arithmetic.
For each DDRS that specifies natural number arithmetic we require that $0$ is a constant 
in its signature and that both
$t+0=t$ and $0\cdot t=0$ hold for each closed term $t$. 

\medskip

All DDRSs further discussed in this paper are based on a signature $\Sigma$ 
that contains the unary minus function $-(\_)$ and will be defined in a pairwise manner:
\begin{enumerate}
\item
a DDRS $(\Sigma\setminus\{-(\_)\}, E_n)$ for natural number arithmetic, and 
\item
a DDRS $(\Sigma, E_z)$, where $E_n\subset E_z$.
\end{enumerate}

\begin{table}
\hrule
\begin{minipage}[t]{0.48\linewidth}\centering
\begin{Lalign}
\tag*{[Nat1]}
\hspace{-22mm}
x + (y + z) &= (x+y) + z
\hspace{-22mm}
\\
\tag*{[Nat2]}
x+y &= y + x\\
\tag*{[Nat3]
}x+0 &= x\\
\label{Nat4}
\tag*{[Nat4]}
\hspace{-22mm}
0\cdot x &= 0
\end{Lalign}
\end{minipage}
\hspace{-8mm}
\begin{minipage}[t]{0.556\linewidth}\centering
\begin{Lalign}
\tag*{[Nat5]}
(x \cdot y) \cdot z &= x \cdot (y \cdot z) \\
\tag*{[Nat6]}
x \cdot y &= y \cdot x\\
\tag*{[Nat7]}
1\cdot x&=x\\
\tag*{[Nat8]}
x \cdot (y + z) & = (x \cdot y) + (x \cdot z)
\end{Lalign}
\end{minipage}\vspace{2mm}
\hrule
\caption{
Axioms for natural number arithmetic}
\label{tab:Nat}
\end{table}

\medskip

In this section we consider a simple form of number representation that is related to 
tallying and establishes a unary numeral system based on the constant 0.
The \emph{unary append} is the one-place (postfix) function
\[\_\apue: \Int \to \Int
\] 
and is an alternative notation for the successor function $S(x)$.
The signature we work in is 
\[\Sigma_U=\{0,-(\_),~\_\apue, +, \cdot\}.\]

For natural numbers, the intended normal forms are 0 for zero, and applications of 
the unary append function that define all successor values: each natural
number $n$ is represented by $n$ applications of the unary append to 0 and can be seen
as representing a sequence of $1$'s of length $n$ having 0 as a single prefix, e.g.
\[(0\apue)\apue\]
is the normal form that represents $2$ and can be abbreviated as $011$.
We name the resulting datatype $\Nat_U$. Clearly, two 
different closed normal forms have distinct values in $\Nat$.

For integers, each minus instance $-t$ of a nonzero normal 
form $t$ in $\Nat_U$ is a normal form over $\Sigma_U$, e.g.
\[-((0\apue)\apue)\]
is the normal form that represents $-2$ and can be abbreviated as $-011$. 
We name the resulting datatype $\Int_U$, which satisfies the property that
two distinct closed normal forms have distinct values in $\Int$.

In Section~\ref{sec:3.1} we introduce DDRSs based on $\Sigma_U$. 
In Section~\ref{sec:3.2} we investigate in what way
the DDRS $\dname2$ for the ring of integers
is related.

\subsection{DDRSs for $\Nat_U$ and $\Int_U$}
\label{sec:3.1}
In the left column of Table~\ref{tab:intuview1} we define the DDRS $\nname1$
for the datatype $\Nat_U$ over the signature $\Sigma_U\setminus\{-(\_)\}$.
With the interpretation rule
\[\llbracket x\apue\rrbracket=\llbracket x\rrbracket+1,\]
it follows that the equations in Table~\ref{tab:intuview1}
are semantic consequences of the axioms for natural number arithmetic in Table~\ref{tab:Nat}. 

The transition to DDRSs for $\Int_U$ can be taken in different ways. 
In Table~\ref{tab:intuview1} we provide the DDRS $\zname1$ that defines 
an extension of $\nname1$ to integer numbers (thus, to the datatype $\Int_U$). 
With the above-mentioned interpretation rule
it follows that the equations in Table~\ref{tab:intuview1}
are semantic consequences of the axioms for commutative rings (Table~\ref{tab:CR}).

\begin{table}
\centering
\hrule
\begin{minipage}[t]{0.49\linewidth}\centering
\begin{Lalign}
\label{U1}
\tag*{\textup{[U1]}}
x+0 &=x
\\
\label{U2}
\tag*{\textup{[U2]}}
x + (y\apue) &= (x\apue)+y
\\[2mm]
\label{U3}
\tag*{\textup{[U3]}}
x \cdot 0 &=0\\
\label{U4}
\tag*{\textup{[U4]}}
x \cdot (y\apue) &= x+(x\cdot y)
\end{Lalign}
\end{minipage}
\hspace{-1mm}
\begin{minipage}[t]{0.50\linewidth}\centering
\begin{Lalign}
\label{U5}
\tag*{\textup{[U5]}}
-0&=0\\
\label{U6}
\tag*{\textup{[U6]}}
(-(x\apue))\apue&=-x\\
\label{U7}
\tag*{\textup{[U7]}}
-(-x) &=x
\\[2mm]
\label{U8}
\tag*{\textup{[U8]}}
x+(-y)&=-((-x)+y)
\\[0mm]
\label{U9}
\tag*{\textup{[U9]}}
x \cdot (-y) &=-(x\cdot y)
\end{Lalign}
\end{minipage}\vspace{4mm}
\hrule
\caption{DDRSs $\nname1$ for~$\Nat_U$ (left column) and $\zname1$ for~$\Int_U$ that 
specify natural number and integer arithmetic}
\label{tab:intuview1}
\end{table}

\begin{thm}
\label{thm:3.1}
The DDRSs $\nname1$ for $\Nat_U$ and $\zname1$
for $\Int_U$ (Table~\ref{tab:intuview1}) are ground-complete.
\end{thm}

\begin{proof}
The AProVE tool~\cite{AProVE} finds that the DDRS $\zname1$ is terminating 
(and therefore the DDRS $\nname1$ is terminating as well).
It remains to be shown that both these DDRSs are ground-confluent. 
We first consider the DDRS  $\zname 1$.
Define the set $N$ as follows:
\begin{align*}
N&=\{0\}\cup N^+\cup N^-,\\
N^+&=\{0\apue\}\cup \{t\apue\mid t\in N^+\},\\
N^-&=\{-t\mid t\in N^+\}.
\end{align*}
It immediately follows that if $t\in N$, then $t$ is a normal form (no rewrite 
rule applies).
In order to prove ground-confluence it suffices to show that
for each closed term $t$ over $\Sigma_U$, either $t\in N$ or $t$ has a 
rewrite step, so that each normal form is in $N$. 
As in the proof of Theorem~\ref{thm:2.2}, this is sufficient because $\zname1$ is terminating, 
all equations of $\zname 1$ are semantic consequences
of the axioms for commutative rings, and distinct closed 
normal forms have distinct values in $\Int$.

We prove this by structural induction on $t$.
The base case is simple: if $t=0$, then  $t\in N$.
For the induction step we have to distinguish four cases:
\begin{enumerate}
\item 
Case $t=-r$. Assume that $r\in N$ and apply case distinction on $r$:
\begin{itemize}
\item
if $r=0$, then $t$ has a rewrite step by equation~\ref{U5},
\item
if $r=r'\apue$, then $t\in N$,
\item
if $r=-(r'\apue)$, then $t$ has a rewrite step by equation~\ref{U7}.
\end{itemize}
\item
Case $t=r\apue$. Assume that $r\in N$ and apply case distinction on $r$:
\begin{itemize}
\item
if $r=0$, then $t\in N$,
\item
if $r=r'\apue$, then $t\in N$,
\item
if $r=-(r'\apue)$, then $t$ has a rewrite step by equation~\ref{U6}.
\end{itemize}
\item 
Case $t=u+r$. Assume that $u,r\in N$ and apply case distinction on $r$:
\begin{itemize}
\item
if $r=0$, then $t$ has a rewrite step by equation~\ref{U1},
\item
if $r=r'\apue$, then $t$ has a rewrite step by equation~\ref{U2},
\item
if $r=-(r'\apue)$, then $t$ has a rewrite step by equation~\ref{U8}.
\end{itemize}
\item 
Case $t=u\cdot r$. Assume that $u,r\in N$ and apply case distinction on $r$:
\begin{itemize}
\item
if $r=0$, then $t$ has a rewrite step by equation~\ref{U3},
\item
if $r=r'\apue$, then $t$ has a rewrite step by equation~\ref{U4},
\item
if $r=-(r'\apue)$, then $t$ has a rewrite step by equation~\ref{U9}.
\end{itemize}
\end{enumerate}

Ground-confluence of the DDRS $\nname 1$ follows in a similar way by restricting 
the above proof to the set
of nonnegative normal forms. Moreover, a confluence proof for the DDRS $\nname 1$ was found 
by the confluence prover CSI~\cite{csi} 
at \url{http://cocoweb.uibk.ac.at/}
(property [{CR}] and options [{2020, TRS, CSI}])
with the input file \url{NAT1.trs} that is available at
\url{https://arxiv.org/src/1608.06212/anc/}.  
\end{proof}

The following example shows that with respect to addition of negative normal forms, 
the DDRS $\zname1$ is not deterministic:

\begin{Ralign}
\label{redu:X}
(-011)+01\stackrel{\text{\ref{U2}}}\longrightarrow((-011)1)+0
\left[\begin{array}{l}
\stackrel{\text{\ref{U6}}}\longrightarrow(-01)+0\stackrel{\text{\ref{U1}}}\longrightarrow 
\\[2mm]
\stackrel{\text{\ref{U1}}}\longrightarrow(-011)1\hfill\stackrel{\text{\ref{U6}}}
\longrightarrow 
\end{array}\right] {-}01.
\\[-3mm]
\nonumber
\end{Ralign}

However, the DDRS $\nname1$ for $\Nat_U$ satisfies the following property.

\begin{prop}
\label{prop:3.2}
With respect to addition and multiplication of closed normal forms, 
the DDRS $\nname1$ in Table~\ref{tab:intuview1} is \textbf{deterministic}, 
that is, for closed normal forms $t,t'$, in each state of rewriting of $t+t'$ 
and $t\cdot t'$ to their normal form, only one equation (rewrite rule) applies.
\end{prop}

\begin{proof}
The case for addition is simple:
applicability of equations~\ref{U1} or~\ref{U2} excludes the other.
\\[4pt]
The case for multiplication follows by structural induction on $t'$.
\\[4pt]
Case $t'=0$. Equation~\ref{U3} defines the only possible rewrite step.
\\[4pt]
Case $t'=r\apue$. Equation~\ref{U4} defines the only possible rewrite step,
resulting in $t+(t\cdot r)$. By induction, $t\cdot r$ rewrites deterministically to 
some normal form $t_n$, say $t\cdot r=t_0$ and $t_0 \to t_1\to\!\!\!\!\to t_n$ for some 
$n>0$. It is easily seen that $t_i$ is the only redex in $t+ t_i$ for $i\leq n$, hence 
also $t+(t\cdot r)$ rewrites deterministically to $t+t_n$. 
By the case for addition, the latter term rewrites deterministically.
\end{proof}

\subsection{From the ring of integers to unary view}
\label{sec:3.2}
In this section we relate the DDRS $\dname2$ for the ring of integers
$\Int_r$ to integer arithmetic as defined in the previous section. 
If we use $t\apue$ as an alternative notation for $t+1$ in $\dname2$ 
and then delete all equations that contain $1$ as a constant 
(thus  $0\apue=1$, $x\cdot 1 = 1$, and $(-1)\apue=0$), we 
obtain the DDRS $\zname2$ given in Table~\ref{tab:int2}, which provides an 
alternative specification of integer arithmetic over the signature $\Sigma_U$ 
comparable to the DDRS $\zname1$ for $\Int_U$ defined in Table~\ref{tab:intuview1}.
Equations~\ref{Um2} (replacing~\ref{U2}) and~\ref{Um4} (replacing~\ref{U4}) are new.
Clearly, $\ref{U1}+\ref{Um2}+\ref{U3}+\ref{Um4}$ define an alternative DDRS $\nname2$
for natural number arithmetic, and these equations are semantic consequences of the 
axioms for natural number arithmetic (Table~\ref{tab:Nat}). 

\begin{table}
\centering
\hrule
\begin{minipage}[t]{0.49\linewidth}\centering
\begin{Lalign}
\tag*{\ref{U1}}
x+0 &=x\\
\label{Um2}
\tag*{\textup{[U2$'$]}}
x + (y\apue) &= (x+y)\apue
\\[2mm]
\tag*{\ref{U3}}
x \cdot 0 &=0\\
\label{Um4}
\tag*{\textup{[U4$'$]}}
x \cdot (y\apue) &=(x\cdot y)+x
\end{Lalign}
\end{minipage}
\begin{minipage}[t]{0.50\linewidth}\centering
\begin{Lalign}
\tag*{\ref{U5}}
-0&=0
\\[0mm]
\tag*{\ref{U6}}
(-(x\apue))\apue&=-x
\\
\tag*{\ref{U7}}
-(-x) &=x
\\[2mm]
\tag*{\ref{U8}}
x+(-y)&=-((-x)+y)
\\[0mm]
\tag*{\ref{U9}}
x \cdot (-y) &=-(x\cdot y)
\end{Lalign}
\end{minipage}\vspace{4mm}
\hrule
\caption{
DDRSs $\nname2$ for~$\Nat_U$ (left column) and $\zname2$ for~$\Int_U$ that 
specify natural number and integer arithmetic}
\label{tab:int2}
\end{table}

\begin{thm}
\label{thm:3.3}
The DDRSs $\nname2$ for $\Nat_U$ and $\zname2$ for $\Int_U$ (Table~\ref{tab:int2}) are 
ground-complete.
\end{thm}

\begin{proof}
The AProVE tool~\cite{AProVE} finds that these DDRSs are terminating, so
it remains to be shown that both these DDRSs are ground-confluent. We first consider 
the DDRS $\zname2$. It immediately
follows from the proof of Theorem~\ref{thm:3.1} that~\ref{Um2} admits a 
rewrite step if~\ref{U2} does, and~\ref{Um4} admits a 
rewrite step if~\ref{U4} does.
Hence the DDRS $\zname2$ is ground-confluent.

By restricting this proof to the set
of nonnegative normal forms, it follows that the DDRS $\nname2$ is ground-confluent.
Moreover, a confluence proof for the DDRS $\nname2$ has been found 
by the confluence prover CSI~\cite{csi}
at \url{http://cocoweb.uibk.ac.at/}
(property [{CR}] and options [{2020, TRS, CSI}])
with the input file \url{NAT2.trs} that is available at
\\
\url{https://arxiv.org/src/1608.06212/anc/}.  
\end{proof}

Furthermore, Proposition~\ref{prop:2.4} and the above proof
imply that the DDRS $\zname2$ 
is deterministic with respect to rewriting a sum of two 
nonnegative closed normal forms.
However, $\zname2$ is not deterministic with respect to multiplication of two 
nonnegative normal forms:

\[0\cdot(0\apue)\stackrel{\text{\ref{Um4}}}\longrightarrow
(0\cdot 0) + 0\left[\begin{array}{l}
\stackrel{\text{\ref{U1}}}\longrightarrow ~0\cdot 0\hfill\stackrel{\text{\ref{U3}}}\longrightarrow\\[2mm]
\stackrel{\text{\ref{U3}}}\longrightarrow 0+0\stackrel{\text{\ref{U1}}}\longrightarrow
\end{array}\right]0.
\]

\section{DDRSs for combining unary, binary and decimal view}
\label{sec:4}
In this section we define various DDRSs for the unary, binary and decimal view.
In Section~\ref{sec:4.1} we fix a signature that comprises these views. In Section~\ref{sec:4.2}
we define
two canonical term algebras that represent the unary view: $\Nat_{ubd}$ and $\Int_{ubd}$.
Their defining DDRSs $\nname{ubd}$ and $\zname{ubd}$, respectively, comprise the conversion from numbers 
in binary or decimal view to unary view (hence the ordering in the subscript).

In Section~\ref{sec:4.3} we define DDRSs for a binary view of natural and integer arithmetic, and in 
Section~\ref{sec:4.4} we do the same for a decimal view of natural and integer arithmetic.
 
\subsection{Digits, a large signature, and two canonical term algebras}
\label{sec:4.1}
Digits are elements of the set $D=\{0, 1, 2, 3, 4, 5, 6, 7, 8, 9\}$, ordered in the common way:
\[0 < 1 < 2 < 3 <  4 < 5 <  6 < 7 < 8 < 9.\]

For the digits $0, 1, \dots, 8$ we denote with $i^{\prime}$ the successor digit of $i$ 
in the given enumeration.
In Table~\ref{fig:enumdig} the successor notation on digits is specified as a transformation 
of syntax, and we adopt this notation throughout the paper.

\begin{table}
\centering
\hrule
\begin{align*}
0^{\prime} &\equiv 1&3^{\prime} &\equiv 4&6^{\prime} &\equiv 7\\
1^{\prime} &\equiv 2&4^{\prime} &\equiv 5&7^{\prime} &\equiv 8\\
2^{\prime} &\equiv 3&5^{\prime} &\equiv 6&8^{\prime} &\equiv 9
\end{align*}
\hrule
\caption{Enumeration and successor notation of digits of type $\Int$}
\label{fig:enumdig}
\end{table}

In forthcoming DDRSs we will often add tags of the form
\[\text{[N$n.i\hspace{.4mm}]_{i=k}^\ell~~t=r$}\]
with $n,k,\ell \in \Nat$  (in ordinary, decimal notation) and $k < \ell$, 
which represents the following $\ell - k+1$ equations:
\[\text{[N$n.k$]~~} t[k/i]=r[k/i],~\dots ,~\text{[N$n.\ell$]~~} t[\ell/i]=r[\ell/i],\]
thus with $i$ instantiated from $k$ to $\ell$. 
Occasionally, we will use this notation with two  ``digit counters'', as in
\[\text{[N$n.i.j\hspace{.4mm}]_{i,j=k}^\ell~~t=r$},
\]
for a concise representation of the following $(\ell-k+1)^2$ equations:
\[\begin{aligned}
&\text{[N$n.k.k$]~~} t[k/i][k/j]=r[k/i][k/j], ~\dots,&\text{[N$n.k.\ell$]~~} 
t[k/i][\ell/j]=r[k/i][\ell/j],\\
&\ldots,\\
&\text{[N$n.\ell.k$]~~} t[\ell/i][k/j]=r[\ell/i][k/j], 
~~\dots,&\text{[N$n.\ell.\ell$]~~} t[\ell/i][\ell/j]=r[\ell/i][\ell/j].~
\end{aligned}\]

The signature $\Sigma_{\Int}$ considered henceforth has the following elements:
\begin{enumerate}

\item a sort $\Int$, 

\item for digits the ten  constants $0,1,2,3,4,5,6,7,8,9$, 

\item three one-place functions 
$S,P,-:\Int \to \Int$,
``successor'', ``predecessor'', and  ``minus'', respectively,

\item addition and multiplication (infix) 
$+,\cdot: \Int \times \Int \to \Int,$

\item 
\label{it:5}
two one-place functions (postfix)
\(\_\apbz, ~\_\apbu: \Int \to \Int,\)
``binary append zero'' and ``binary append one'',
these functions will be used for binary 
notation,

\item 
\label{it:6}
ten one-place functions (postfix)
\begin{align*}
&\_\apd 0, ~\_\apd 1, ~\_\apd 2, ~\_\apd 3, ~\_\apd 4, ~\_\apd 5,
~\_\apd 6, ~\_\apd 7, ~\_\apd 8, ~\_\apd 9:\Int \to \Int,
\end{align*}
``decimal append zero'', ...,``decimal append nine'', to be used for decimal 
notation.
\end{enumerate}

For the unary view, the normal forms we will consider are the constant (digit) 0 and
the classical successor terms, that is 
\[0,S(0), S(S(0)),...\]
and all minus instances $-t$ of each nonzero normal form $t$, e.g.~$-(S(S(0)))$.
We shall use the following abbreviation, where $i$ is a digit:
$S^i(t)$ stands for $i$ applications of the successor function $S$ to $t$, 
thus $S^0(t)=t$ and 
$S^{i'}(t)=S(S^i(t))$.

For the binary view and for the decimal view, we provide one DDRS for each.
Normal forms are all appropriate digits, all applications of the respective 
append functions to a nonzero normal form, and all minus instances $-t$ of each 
such normal form $t$ that differs from 0. For example,
\[(9\apd7)\apd5\quad \text{and}\quad ((1\apbz)\apbu)\apbu\]
represent the decimal number 975 
(with the interpretation rule $\llbracket x\apd i\rrbracket=10\cdot \llbracket x\rrbracket+i$),
and the binary number $1011$, respectively
(with the interpretation rule $\llbracket x\apb i\rrbracket=2\cdot \llbracket x\rrbracket+i$).

Finally, we will include in forthcoming DDRSs  equations for conversion from the one view to the other.

\subsection{Unary view}
\label{sec:4.2}
Replacing the $t\apue$-occurrences by $S(t)$ in the DDRS $\nname2$ 
defined in Table~\ref{tab:int2} results in equations \ref{S1} -- \ref{S4}  
in Table~\ref{tab:natunview}, which hence define 
a DDRS for natural number 
arithmetic in unary view with successor function. 
We name this DDRS $\nname{ubd}$ and its CTA $\Nat_{ubd}$.
The equations~\ref{Sb} -- \ref{Sd2} define the conversion of terms that employ constructors
from binary or decimal view to unary view.

Integer arithmetic is obtained by replacing 
all $t\apue$-occurrences by $S(t)$ in the DDRS $\zname2$ 
defined in Table~\ref{tab:int2}. 
We name the resuting DDRS $\zname{ubd}$ and its CTA $\Int_{ubd}$.
The proof of Theorem~\ref{thm:3.3} implies the following result.

\begin{table}
\centering
\hrule
\begin{minipage}[t]{0.55\linewidth}\centering
\begin{Lalign}
\label{S1}
\tag*{[S1]}
x+0 &=x\\
\label{S2}
\tag*{[S2]}
x + S(y) &= S(x+y)
\\[2mm]
\label{S3}
\tag*{[S3]}
x \cdot 0 &=0\\
\label{S4}
\tag*{[S4]}
x \cdot S(y) &= (x\cdot y)+x
\\[6mm]
\label{Sb}
\tag*{[\textbf{Sb.}$i\,]_{i=0}^1$}
x \apb i  &= (x\cdot S(1)) + i\\
\label{Sd1}
\tag*{[\textbf{Sd1.}$i\hspace{.4mm}]_{i=0}^8$}
i^{\prime} &= S(i)\hspace{-3cm}\\
\label{Sd2}
\tag*{[\textbf{Sd2.}$i\hspace{.4mm}]_{i=0}^9$}
x \apd i &= (x \cdot S(9)) + i
\end{Lalign}
\end{minipage}
\hspace{-9mm}
\begin{minipage}[t]{0.49\linewidth}\centering
\begin{Lalign}
\label{S5}
\tag*{[S5]}
-0&=0\\
\label{S6}
\tag*{[S6]}
S(-S(x))&=-x\\
\label{S7}
\tag*{[S7]}
-(-x) &=x
\\[2mm]
\label{S8}
\tag*{[S8]}
x+(-y)&=-((-x)+y)
\\[0mm]
\label{S9}
\tag*{[S9]}
x \cdot (-y) &=-(x\cdot y)
\end{Lalign}
\end{minipage}\vspace{4mm}
\hrule
\caption{DDRSs $\nname{ubd}$ for~$\Nat_{ubd}$ (left column)
and $\zname{ubd}$ for~$\Int_{ubd}$}
\label{tab:natunview}
\end{table}

\begin{thm}
\label{thm:natun}  
The DDRSs $\nname{ubd}$ and $\zname{ubd}$ (Table~\ref{tab:natunview}) are ground-complete.
\end{thm}
\begin{proof}
The AProVE tool~\cite{AProVE} finds that these DDRSs are terminating.
We first prove  ground-confluence for the  DDRS $\zname{ubd}$.
It is sufficient to consider the proof of Theorem~\ref{thm:3.3}:
the renaming to successor terms is not relevant, and the `new' equations~\ref{Sb} -- \ref{Sd2} 
rewrite to successor terms and thus preserve ground-confluence.

In a similar way it follows that the DDRS $\nname{ubd}$ is ground-confluent.
Moreover, a ground-confluence proof for the DDRS $\nname{ubd}$ was found by the 
ground-confluence prover AGCP~\cite{ATK}
at \url{http://cocoweb.uibk.ac.at/}
(property [{GCR}] and options [{2020, TRS, AGCP}])
with  input 
\url{https://arxiv.org/src/1608.06212/anc/NATubd.trs}.
\end{proof}

The equations~\ref{S1} -- \ref{S4} that define natural number arithmetic with 
0 and successor function are very common (see, e.g.~\cite{KV03,UK03,WZ95}). 
Note that the DDRS  $\nname{ubd}$ is deterministic with respect to addition of two normal forms
(cf.~Proposition~\ref{prop:2.4}), 
but not with respect to their multiplication (cf. counterexample~\eqref{redu:X} 
in Section~\ref{sec:3.2}).

\subsection{Binary view}
\label{sec:4.3}
In the left column of Table~\ref{tab:binview} we define the DDRS $\nname{bud}$ for a binary 
view of natural numbers that employs the successor function as an auxiliary function.
Leading zeros except for the zero itself are removed by~\ref{b1}, and
successor terms are rewritten according to \ref{b2} -- \ref{b5}.

\begin{table}
\hrule
\begin{minipage}[t]{0.54\linewidth}\centering
\begin{Lalign}
\label{b1}
\tag*{[b1.$i\hspace{.4mm}]_{i=0}^1$}
0\apb i &=i\\[2mm]
\label{b2}
\tag*{[b2]}
S(0) &= 1\\
\label{b3}
\tag*{[b3]}
S(1) &= 1\apbz\\
\label{b4}
\tag*{[b4]}
S(x\apbz) &=x\apbo\\
\label{b5}
\tag*{[b5]}
S(x\apbo) &=S(x)\apbz
\\[2mm]
\label{b6}
\tag*{[b6]}
x+0 &=x\\
\label{b7}
\tag*{[b7]}
x+1 &=S(x)\\
\label{b8}
\tag*{[b8]}
x + (y\apb 0) &= (x+y)+y
\\
\label{b9}
\tag*{[b9]}
x + (y\apb 1) &=S(x + (y\apb 0))
\\[2mm]
\label{b10}
\tag*{[b10]}
x \cdot 0 &=0\\
\label{b11}
\tag*{[b11]}
x \cdot 1 &=x\\
\label{b12}
\tag*{[b12.$i\hspace{.4mm}]_{i=0}^1$}
\\\nonumber
x\cdot(y\apb i) &=((x\cdot y)\apbz )+(x\cdot i)
\\[8mm]
\label{bd1}
\tag*{[\textbf{bd1.$i\hspace{.4mm}]_{i=1}^8$}}
i' &= S(i)
\\
\label{bd2}
\tag*{[\textbf{bd2.$i\hspace{.4mm}]_{i=0}^9$}}
x\apd  i &= (x \cdot  S(9)) + i
\end{Lalign}
\end{minipage}
\hspace{-12mm}
\begin{minipage}[t]{0.51\linewidth}\centering
\begin{Lalign}
\label{b13}
\tag*{[b13]}
-0 &=0\\
\label{b14}
\tag*{[b14]}
-(-x) &= x
\\[2mm]
\label{b15}
\tag*{[b15]}
P(0) & =-1\\
\label{b16}
\tag*{[b16]}
P(1) &= 0\\
\label{b17}
\tag*{[b17]}
P(x \apbz) &= P(x)\apbu\\
\label{b18}
\tag*{[b18]}
P(x \apbo) &= x\apbz\\
\label{b19}
\tag*{[b19]}
P(-x) & = -S(x)
\\[2mm]
\label{b20}
\tag*{[b20]}
S(-1) &=0\\
\label{b21}
\tag*{[b21]}
S(-( x\apbz)) &= -(P(x) \apbu) \\
\label{b22}
\tag*{[b22]}
S(-(x \apbu)) &= -(x \apbz)
\\[2mm]
\label{b23}
\tag*{[b23]}
(-x) \apbz &= -(x \apbz)\\
\label{b24}
\tag*{[b24]}
(-x) \apbu &= -(P(x) \apbu)
\\[2mm]
\label{b25}
\tag*{[b25]}
x + (-y) &= -((-x)+y)\\[2mm]
\label{b26}
\tag*{[b26]}
x \cdot (-y) &= -(x \cdot y)
\end{Lalign}
\end{minipage}
\vspace{4mm}
\hrule
\caption{DDRSs $\nname{{bud}}$ for~$\Nat_{bud}$ (left column) and $\zname{{bud}}$ for~$\Int_{bud}$}
\label{tab:binview}
\end{table}

\begin{thm}
The DDRS $\nname{bud}$ for $\Nat_{bud}$ (Table~\ref{tab:binview}) is ground-complete.
\end{thm}

\begin{proof}
The AProVE tool~\cite{AProVE} finds that this DDRS is terminating.
Ground-confluence follows as in the proof in Appendix~\ref{app:A.1} restricted 
to the set of nonnegative normal forms.
\end{proof}

In the right column of Table~\ref{tab:binview} minus and predecessor are introduced 
and the transition from a signature
for natural numbers to a signature for integers is made; the rules in this table 
define the DDRS $\zname{bud}$ and the canonical term algebra $\Int_{bud}$ that is 
isomorphic to the canonical term algebra 
$\Int_{ubd}$ of the DDRS $\zname{ubd}$ in Table~\ref{tab:natunview}.
The DDRS $\zname{bud}$ contains 
twenty-eight (non-parametric) equations for the specification of numbers, addition, 
and multiplication. 
A brief comment on equations~\ref{b23} and \ref{b24}: 
\[(-x)\apb i\]
should be equal to 
$(-(x\apb 0)) + i,$
so 
$(-x)\apbz=-( x\apbz)$, and
$(-x) \apbu$ is determined by
\[-(P( x\apbz))\stackrel{\text{\footnotesize\ref{b24}}}=-(P(x) \apbu).\]
Equations~\ref{b21} and \ref{b22} can be explained in a similar way:
\begin{align*}
S(-( x\apbz))&
\text{ should be equal to } 
-(P(x\apbz))=-(P(x)\apbu),\\
S(-( x\apbu))&
\text{ should be equal to } 
-(P(x\apbu))=-(x\apbz).
\end{align*}
Normal forms for $\Int_{bud}$
are $0$, $1$, all applications of $\_\apbz$ and $\_\apbu$
to a nonzero normal form, and all minus instances $-t$ of each 
such normal form $t$ that differs from 0.

Note that the equations in Table~\ref{tab:binview} are semantic 
consequences of the axioms for commutative rings (Table~\ref{tab:CR}),
and that two distinct closed normal forms have distinct values in~$\Int$.

\begin{thm}
The DDRS $\zname{bud}$ for $\Int_{bud}$ (Table~\ref{tab:binview}) is ground-complete.
\end{thm}
\begin{proof}
The AProVE tool~\cite{AProVE} finds that the DDRS $\zname{bud}$ is  
terminating. In Appendix~\ref{app:A.1} we prove that the DDRS  $\zname{dub}$ 
is ground-confluent.
\end{proof}

\subsection{Decimal view}
\label{sec:4.4}
We provide DDRSs for the decimal view that are straightforward generalizations of 
$\nname{bud}$ and $\zname{bud}$ to the decimal view.
In the left column of Table~\ref{tab:indecview} we define the 
DDRS $\nname{dub}$ for decimal 
natural numbers that defines the canonical term algebra 
$\Nat_{dub}$, the datatype in which unary and binary view are 
derived representations. This DDRS consists of twelve (parametric) equations,
and another one for conversion from binary view
(72 equations in total). 
The datatype $\Nat_{dub}$  is isomorphic to the canonical term  algebra $\Nat_{ubd}$ of 
the DDRS $\nname{ubd}$ in Table~\ref{tab:natunview}. 
Leading zeros except for the zero itself are removed by \ref{d1}, and
successor terms are rewritten according to \ref{d2} -- \ref{d5}.
In equation~\ref{d8}, the notation $\boldsymbol{+}^{\!10}$ is used for a 
nested sum:
\[x\boldsymbol{+}^{\!1} y=x+y\quad\text{and for $n=1,...,9$,}\quad 
x\boldsymbol{+}^{\!n{+}1}y=(x\boldsymbol{+}^{\!n} y)+y.
\]
Rewriting from binary notation is part of this DDRS, and the equation scheme~\ref{db1}
serves that purpose. 
Clearly, two distinct closed normal forms have distinct values in $\Nat$, and
all equations in $\nname{dub}$ are semantic consequences of the axioms for 
natural number arithmetic (Table~\ref{tab:Nat}).

\begin{thm}
\label{thm:4.4}
The DDRS $\nname{dub}$ for $\Nat_{dub}$ (Table~\ref{tab:indecview}) is ground-complete.
\end{thm}

\begin{proof}
The AProVE tool~\cite{AProVE} finds that this DDRS is terminating.
Ground-confluence follows as in the proof in Appendix~\ref{app:A.2}, restricted 
to the set of nonnegative normal forms.
\end{proof}

Before we extend the DDRS $\nname{dub}$ to the integers, we define 
 ``10's complement'', notation $i^\star$, for 
digits $i\in\{1,...,9\}$ in Table~\ref{fig:subdig}, which can be characterised
by the equation scheme
\[i^\star=10-i.\]

\begin{table}
\centering
\hrule
\begin{align*}
1^{\star} &\equiv 9&4^{\star} &\equiv 6&7^{\star} &\equiv 3\\
2^{\star} &\equiv 8&5^{\star} &\equiv 5&8^{\star} &\equiv 2\\
3^{\star} &\equiv 7&6^{\star} &\equiv 4&9^{\star} &\equiv 1
\end{align*}
\hrule
\caption{10's complement notation for decimal digits}
\label{fig:subdig}
\end{table}

In Table~\ref{tab:indecview}, we define the DDRS $\zname{dub}$ in which minus and predecessor are added.
In rule scheme~\ref{d24} we employ the notation $i^\star$.
The canonical term algebra thus defined is named $\Int_{dub}$ and is isomorphic to \
$\Int_{ubd}$ of the specification in Table~\ref{tab:natunview}. The DDRS $\zname{dub}$
contains  126 equations in total, including two for conversion from binary view.

\begin{table}
\hrule
\begin{minipage}[t]{0.54\linewidth}\centering
\begin{Lalign}
\label{d1}
\tag*{[d1.$i\hspace{.4mm}]_{i=0}^9$}
0 \apd i &= i\\[2mm]
\label{d2}
\tag*{[d2.$i\hspace{.4mm}]_{i=0}^8$}
S(i) &=i^{\prime}\\
\label{d3}
\tag*{[d3]}
S(9) &= 1 \apd 0\\
\label{d4}
\tag*{[d4.$i\hspace{.4mm}]_{i=0}^8$}
S(x \apd i) &= x \apd i^{\prime}\\
\label{d5}
\tag*{[d5]}
S(x \apd 9) &= S(x) \apd 0\\[2mm]
\label{d6}
\tag*{[d6]}
x + 0 &=x\\
\label{d7}
\tag*{[d7.$i\hspace{.4mm}]_{i=1}^9$}
x+i &= S^i(x)\\
\label{d8}
\tag*{[d8]}
x+(y \apd 0) &= x \boldsymbol{+}^{\!10}y
\\
\label{d9}
\tag*{[d9.$i\hspace{.4mm}]_{i=1}^9$}
\\\nonumber
x+(y \apd i) &= S^i(x+(y \apd 0))
\\[2mm]
\label{d10}
\tag*{[d10]}
x \cdot 0 &=0\\
\label{d11}
\tag*{[d11.$i\hspace{.4mm}]_{i=0}^8$}
x \cdot i^{\prime}  & = (x \cdot i) + x\\
\label{d12}
\tag*{[d12.$i\hspace{.4mm}]_{i=0}^9$}
\\\nonumber
x \cdot (y \apd i) &= ((x \cdot y) \apd 0) + (x \cdot i)
\\[6mm]
\label{db1}
\tag*{[\textbf{db1.$i\hspace{.4mm}]_{i=0}^1$}}
x\apb  i &= (x +x) + i
\end{Lalign}
\end{minipage}
\hspace{-18mm}
\begin{minipage}[t]{0.56\linewidth}\centering
\begin{Lalign}
\label{d13}
\tag*{[d13]}
-0 &=0\\
\label{d14}
\tag*{[d14]}
-(-x) &= x
\\[3mm]
\label{d15}
\tag*{[d15]}
P(0)&=-1\\
\label{d16}
\tag*{[d16.$i\hspace{.4mm}]_{i=0}^8$}
P(i^\prime)  &=  i\\
\label{d17}
\tag*{[d17]}
P(x \apd 0) &= P(x) \apd 9
\\
\label{d18}
\tag*{[d18.$i\hspace{.4mm}]_{i=0}^8$}
P(x \apd i^\prime) &= x \apd i\\
\label{d19}
\tag*{[d19]}
P(-x) &= -S(x)
\\[3mm]
\label{d20}
\tag*{[d20.$i\hspace{.4mm}]_{i=0}^8$}
S(-i^\prime)  &= - i\\
\label{d21}
\tag*{[d21]}
S(-(x \apd 0)) &= -(P(x) \apd 9)
\\
\label{d22}
\tag*{[d22.$i\hspace{.4mm}]_{i=0}^8$}
\\\nonumber
S(-(x \apd i^\prime)) &= -(x \apd i)
\\[2mm]
\label{d23}
\tag*{[d23]}
(-x) \apd 0 &= -(x \apd 0)\\[0mm]
\label{d24}
\tag*{[d24.$i\hspace{.4mm}]_{i=1}^9$}
(-x) \apd i &= -(P(x) \apd i^{\star})
\hspace{-18mm}
\\[4mm]
\label{d25}
\tag*{[d25]}
x + (-y) &= -((-x)+y)\\[2mm]
\label{d26}
\tag*{[d26]}
x \cdot (-y) &= -(x \cdot y)
\end{Lalign}
\end{minipage}
\vspace{4mm}
\hrule
\caption{DDRSs $\nname{dub}$ for~$\Nat_{dub}$ (left column) and $\zname{dub}$ 
for~$\Int_{dub}$ that specify natural number and integer arithmetic in decimal view, 
employing $i^\star$ from Table~\ref{fig:subdig}}
\label{tab:indecview}
\end{table}

The (twenty) equations captured by \ref{d21} -- \ref{d24} can be explained
in a similar fashion as was done in the previous section for \ref{b21} -- \ref{b24},
for example, 
\[(-5)\apd3\]
should be equal to $-(5\apd0)+3=-(4\apd 7)$, and this follows immediately from  
equation~[d24.3].

The equations of the DDRS $\zname{dub}$ are semantic 
consequences of the equations for commutative rings  (Table~\ref{tab:CR}).
It is also clear that two distinct closed normal forms have distinct values in $\Int$.

\begin{thm}
\label{thm:4.5}
The DDRS $\zname{dub}$ for $\Int_{dub}$ (Table~\ref{tab:indecview}) is ground-complete.
\end{thm}

\begin{proof}
The AProVE tool~\cite{AProVE} finds that this DDRS is terminating.
In Appendix~\ref{app:A.2} we prove that the DDRS  $\zname{dub}$ is ground-confluent.
\end{proof}

\section{DDRSs with digit tree constructors}
\label{sec:5}

Having defined DDRSs that employ (postfix) digit append functions in Sections~\ref{sec:3} 
and~\ref{sec:4},
we now consider the more general \emph{digit tree constructor} functions. 
For the binary view, this approach is followed by \textsc{Bouma \& Walters} in~\cite{BW89}; 
for a view based on any radix (number base), this approach is further continued in 
\textsc{Walters}~\cite{Wal94} and \textsc{Walters \& Zantema}~\cite{WZ95}, where the constructor 
is called \emph{juxtaposition} 
because it goes with the absence of a function symbol in order to be close to ordinary decimal 
and binary notation.  

We extend the signature $\Sigma_\Int$ defined in Section~\ref{sec:4.1} with the following three 
functions (infix):
\[\concu,\concb, \concd: \Int\times \Int \to \Int,\]
called ``unary digit tree constructor function'', ``binary digit tree constructor function'',
and ``decimal digit tree constructor function'',
and to be used for unary, binary notation and decimal notation, respectively.
The latter two constructors serve to represent 
positional notation and satisfy the semantic equations 
\[\text{$\llbracket x\concb y\rrbracket=2\cdot\llbracket x\rrbracket+\llbracket y\rrbracket$\quad
and 
\quad$\llbracket x\concd y\rrbracket=10\cdot\llbracket x\rrbracket+\llbracket y\rrbracket$.}\]

For integer numbers in decimal view or binary view, normal forms are the relevant digits, 
all applications of the respective constructor with left argument a nonzero normal form and 
right argument a digit, and all minus instances $-t$ of each such nonzero normal form $t$, 
these satisfy $\llbracket-(t)\rrbracket=-(\llbracket t\rrbracket)$.
E.g.,
\[(9\concd7)\concd5\quad \text{and}\quad ((1\concb0)\concb 1)\concb 1\]
represent the decimal number 975
and the binary number $1011$, respectively, and the normal form that represents the additional 
inverse of the latter is $-(((1\concb0)\concb1)\concb1)$.
A minor complication with decimal and binary digit tree constructors is that we now have to
consider rewritings such as
\[2\concd (1\concd 5)=(2+1)\concd 5=3\concd 5 \quad(=35),\]
which perhaps are somewhat non-intuitive. For integers in unary view, thus with 
unary digit tree constructor, this complication is absent (see Section~\ref{sec:5.1}).

We keep the presentation of the resulting DDRSs (those defining the binary and decimal view 
are based on \textsc{Walters}~\cite{Wal94} and \textsc{Walters \& Zantema}~\cite{WZ95}) 
minimal in the sense that equations for conversion from the one view
to the other are left out. Of course, it is easy to define such equations.
Also, equations for conversion to and from the datatypes defined in Section~\ref{sec:4} are 
omitted, although such equations are also easy to define.
 
\subsection{Unary view with digit tree constructor}
\label{sec:5.1}
For naturals in this particular unary view, normal forms are 0 and expressions 
$t\concu 0$ 
with $t$ a normal form (thus, with association of $\concu$ to the left).
Of course, the phenomenon of ``removing leading zeros'' does not exist in this unary 
view. The resulting datatype $\Nat_{ut}$ is defined by the DDRS $\nname{ut}$
in Table~\ref{tab:intuaview}.

\begin{table}
\centering
\hrule
\begin{minipage}[t]{0.44\linewidth}\centering
\begin{Lalign}
\label{ut1}
\tag*{[ut1]}
x\concu(y\concu z)&=(x\concu y)\concu z\\[2mm]
\label{ut2}
\tag*{[ut2]}
x+0 &=x\\
\label{ut3}
\tag*{[ut3]}
x + (y\concu 0)&= (x+y)\concu 0
\hspace{-12mm}
\\[2mm]
\label{ut4}
\tag*{[ut4]}
x \cdot 0 &=0\\
\label{ut5}
\tag*{[ut5]}
x \cdot (y\concu 0)&= (x\cdot y)+x
\end{Lalign}
\end{minipage}
\hspace{-12mm}
\begin{minipage}[t]{0.58\linewidth}\centering
\begin{Lalign}
\label{ut6}
 \tag*{[ut6]}
-0&=0\\
\label{ut7}
\tag*{[ut7]}
-(-x) &=x\\[2mm]
\label{ut8}
\tag*{[ut8]}
0 \concu (-(x\concu 0)) &= -x
\\
\label{ut9}
\tag*{[ut9]}
(x\concu 0) \concu (-(y\concu 0)) &= x\concu (-y)
\\
\label{ut10}
\tag*{[ut10]}
(-(x\concu 0)) \concu 0 &= -x\\
\label{ut11}
\tag*{[ut11]}
(-(x\concu 0))\concu(y\concu 0)&= (-x)\concu y\\
\label{ut12}
\tag*{[ut12]}
(-(x\concu 0)) \concu (-(y\concu 0)) &= -((x+y)\concu 0)
\hspace{-18mm}
\\[2mm]
\label{ut13}
\tag*{[ut13]}
x + (-y) &=-((-x)+ y)
\hspace{-18mm}
\\[2mm]
\label{ut14}
\tag*{[ut14]}
x \cdot (-y) &=-(x\cdot y)
\end{Lalign}
\end{minipage}\vspace{4mm}
\hrule
\caption{DDRSs $\nname{ut}$ for~$\Nat_{ut}$ (left column) and $\zname{ut}$ 
for~$\Int_{ut}$ that specify natural number and 
integer arithmetic in unary view with unary digit tree constructor}
\label{tab:intuaview}
\end{table}

The constructor $\concu$ is an associative operator, as is clear from rule~\ref{ut1} 
(in contrast to digit tree constructors for the binary and decimal case).
Moreover, the commutative variants $t\concu r$ and $r\concu t$ rewrite to the same 
normal form, which also is implied by the semantics for closed terms:
\begin{align*}
\llbracket 0 \rrbracket&=0,&\llbracket x+ y \rrbracket&=\llbracket x \rrbracket + 
\llbracket y \rrbracket,\\
\llbracket x\concu y \rrbracket&=\llbracket x \rrbracket + \llbracket y \rrbracket +1,
&\llbracket x\cdot y 
\rrbracket&=\llbracket x \rrbracket \cdot \llbracket y \rrbracket .
\end{align*}
Clearly, all equations in $\nname{ut}$ are semantic consequences of the axioms for 
natural number arithmetic (Table~\ref{tab:Nat}) and
two distinct closed normal forms have distinct values in $\Nat$.

\medskip

The extension to integer numbers can be done in a similar fashion as in the previous section,
thus obtaining normal forms of the form $-(t)$ with $t$ a nonzero normal form in $\Nat_{ut}$.
However, also terms of the form $x\concu (-y)$ and variations thereof have to be considered.
We define the DDRS $\zname{ut}$ in Table~\ref{tab:intuaview} 
and we call the resulting datatype $\Int_{ut}$.
Adding the interpretation rule
$\llbracket -x \rrbracket=-\llbracket x  \rrbracket$, 
it can be easily checked that also $\text{\ref{ut6}}-\text{\ref{ut14}}$ are semantic 
consequences of the axioms for commutative rings,
and that two distinct closed normal forms have distinct values in $\Int$.

\begin{thm}
\label{thm:5.1}
The DDRSs $\nname{ut}$ and $\zname{ut}$ (Table~\ref{tab:intuaview}) are 
ground-complete.
\end{thm}
\begin{proof}
The AProVE tool~\cite{AProVE} finds that both these DDRSs are terminating.
In Appendix~\ref{app:B.1} we prove that the DDRS $\zname{ut}$ is ground-confluent.
 
Ground-confluence of the DDRS $\nname{ut}$ easily follows by restricting this proof 
to the set of nonnegative normal forms.
Moreover, a ground-confluence proof for this DDRS was found by the 
ground-confluence prover AGCP~\cite{ATK}
at \url{http://cocoweb.uibk.ac.at/}
(property [{GCR}], options [{2020, TRS, AGCP}])
with the input  file \url{NATut.trs} available at
\url{https://arxiv.org/src/1608.06212/anc/}.
\end{proof}

\subsection{Binary view with digit tree constructor}
\label{sec:5.2}
For naturals in binary view with the binary digit tree constructor, the associated 
datatype $\Nat_{bt}$ is defined by the DDRS $\nname{bt}$ in Table~\ref{tab:intbview} 
(in the left column). In~\cite{KW16} it is proven that the associated TRS is terminating.

\begin{table}[t]
\hrule
\begin{minipage}[t]{0.48\linewidth}\centering
\begin{Lalign}
\label{bt1}
\tag*{[bt1]}
0\concb x&=x\\
\label{bt2}
\tag*{[bt2]}
x\concb(y\concb z)&=(x+ y)\concb z\\[2mm]
\label{bi3}
\tag*{[bt3]}
0+x &=x\\
\label{bi4}
\tag*{[bt4]}
x+0 &=x\\
\label{bi5}
\tag*{[bt5]}
1+1 &=1\concb 0\\
\label{bi6}
\tag*{[bt6]}
x + (y\concb z) &= y\concb(x+z)\\
\label{bi7}
\tag*{[bt7]}
(x \concb y)+z &= x\concb(y+z)
\\[2mm]
\label{bi8}
\tag*{[bt8]}
x \cdot 0 &=0\\
\label{bi9}
\tag*{[bt9]}
0\cdot x&=0\\
\label{bt10}
\tag*{[bt10]}
1\cdot 1&=1\\
\label{bt11}
\tag*{[bt11]}
x \cdot (y\concb z) &=(x \cdot y)\concb(x\cdot z)
\hspace{-12mm}\\
\label{bt12}
\tag*{[bt12]}
(x \concb y)\cdot z &=(x \cdot z)\concb(y\cdot z)
\hspace{-10mm}
\end{Lalign}
\end{minipage}\vspace{4mm}
\hspace{-10mm}
\begin{minipage}[t]{0.54\linewidth}\centering
\begin{Lalign}
\label{bt13}
\tag*{[bt13]}
-0&=0\\
\label{bt14}
\tag*{[bt14]}
-(-x)&=x\\[2mm]
\label{bt15}
\tag*{[bt15]}
1\concb (-1)&=1\\
\label{bt16}
\tag*{[bt16]}
(x\concb 0)\concb (-1)&=(x\concb(-1))\concb1\\
\label{bt17}
\tag*{[bt17]}
(x\concb 1)\concb (-1)&=(x\concb0)\concb 1\\
\label{bt18}
\tag*{[bt18]}
x\concb (-(y\concb z))&=-((y+(-x))\concb z)
\hspace{-12mm}
\\
\label{bt19}
\tag*{[bt19]}
(-x)\concb y&=-(x\concb(-y))\\[2mm]
\label{bt20}
\tag*{[bt20]}
1+(-1)&=0\\
\label{bt21}
\tag*{[bt21]}
(-1)+1&=0\\
\label{bt22}
\tag*{[bt22]}
(-1)+(-1)&=-(1\concb 0)\\
\label{bt23}
\tag*{[bt23]}
x+(-(y\concb z))&=-(y\concb (z+(-x)))
\hspace{-12mm}
\\
\label{bt24}
\tag*{[bt24]}
(-(x\concb y))+z&=-(x\concb (y+(-z)))
\hspace{-12mm}
\\[2mm]
\label{bt25}
\tag*{[bt25]}
x\cdot (-y)&=-(x\cdot y)\\
\label{bt26}
\tag*{[bt26]}
(-x)\cdot y&=-(x\cdot y)
\end{Lalign}
\end{minipage}
\hrule
\caption{DDRSs $\nname{bt}$ for~$\Nat_{bt}$ (left column) and $\zname{bt}$ 
for~$\Int_{bt}$ for natural number and integer arithmetic in binary view with binary digit 
tree constructor}
\label{tab:intbview}
\end{table}

\medskip

In~\cite{WZ95}, \textsc{Walters \& Zantema} provide a rewriting system for integer arithmetic 
with next to juxtaposition 
and minus, also addition, subtraction and multiplication, and prove its 
ground-completeness with respect to any radix (number base). 
In Table~\ref{tab:intbview} we define the DDRS $\zname{bt}$ that defines the datatype 
$\Int_{bt}$ as a variant of this rewriting system without subtraction (using the 
binary digit tree constructor). 
Clearly, two distinct closed normal forms have distinct values in $\Int$.
Furthermore, the equations in Table~\ref{tab:intbview} are semantic 
consequences of the axioms for commutative rings  (Table~\ref{tab:CR}).

\begin{thm}
\label{thm:5.2}
The DDRSs $\nname{bt}$ and $\zname{bt}$ (Table~\ref{tab:intbview}) are 
ground-complete.
\end{thm}

\begin{proof}
The AProVE tool~\cite{AProVE} finds that both these DDRSs are terminating 
(as reported in~\cite{KW16}).
In Appendix~\ref{app:B.2} we prove that the DDRS $\zname{bt}$ is ground-confluent. 

Ground-confluence of $\nname{bt}$ follows in a similar way by restricting this 
proof to the set of nonnegative normal forms.
\end{proof}

\subsection{Decimal view with digit tree constructor}
\label{sec:5.3}
For the specification of naturals in decimal view with the decimal digit tree 
constructor we make use of successor terms in order to avoid (non-parametric) 
equations such as 
\[\begin{aligned}
&1+1=2,~\dots,&9+8=1\concd 7,&&9+9 = 1\concd 8,\\ 
&\dots,\\
&1\cdot 1=1,~~\dots,&8\cdot 9=7\concd 2,&&~9\cdot 9=8\concd1.~\end{aligned}\]
The associated datatype $\Nat_{dt}$ is defined by the DDRS $\nname{dt}$ in 
Table~\ref{tab:intdecview} (left column). In equations~\ref{dt10} we use 
the notation
\[\textstyle\sumi^i x\]
for $i-1$  repeated applications of $+$ with association to the left, thus
\[\textstyle\sumi^1 x=x \quad\text{and for $i=1,...,8$}, 
\quad\textstyle\sumi^{i+1}x=(\sumi^i x)+x.\]

\begin{table}
\hrule
\begin{minipage}[t]{0.46\linewidth}\centering
\begin{Lalign}
\label{dt1}
\tag*{[dt1]}
0\concd x&=x
\\
\label{dt2}
\tag*{[dt2.$i\hspace{.4mm}]_{i=0}^9$}
\\\nonumber
x\concd (y\concd i)&=(x+y)\concd i
\\[3mm]
\label{dt3}
\tag*{[dt3.$i\hspace{.4mm}]_{i=0}^8$}
S(i)&=i'
\\
\label{dt4}
\tag*{[dt4]}
S(9)&=1\concd 0
\\
\label{dt5}
\tag*{[dt5.$i\hspace{.4mm}]_{i=0}^8$}
S(x\concd i)&=x\concd i'
\\
\label{dt6}
\tag*{[dt6]}
S(x\concd 9)&=S(x)\concd 0
\\[3mm]
\label{dt7}
\tag*{[dt7.$i\hspace{.4mm}]_{i=0}^9$}
x+i&=S^i(x)
\\
\label{dt8}
\tag*{[dt8.$i\hspace{.4mm}]_{i=0}^9$}
\\\nonumber
x+(y\concd i)&=S^i(y\concd x)
\\[3mm]
\label{dt9}
\tag*{[dt9]}
x\cdot 0 &=0\\
\label{dt10}
\tag*{[dt10.$i\hspace{.4mm}]_{i=1}^9$}
x\cdot i&=\textstyle\sumi^i x\\
\label{dt11}
\tag*{[dt11.$i\hspace{.4mm}]_{i=0}^9$}
\\[1mm]\nonumber
x\cdot (y\concd  i)=((x\cdot y)\concd  0)+(x\cdot i)
\hspace{-28mm}
\end{Lalign}
\end{minipage}\vspace{4mm}
\hspace{-16mm}
\begin{minipage}[t]{0.63\linewidth}\centering
\begin{Lalign}
\label{dt12}
\tag*{[dt12]}
-0&=0
\\
\label{dt13}
\tag*{[dt13]}
-(-x)&=x
\\[2mm]
\label{dt14}
\tag*{[dt14.$i\hspace{.4mm}]_{i=0}^8$}
S(-i')&=-i
\\
\label{dt15}
\tag*{[dt15]}
S(-(x\concd 0))&=-(x\concd (-1))
\\
\label{dt16}
\tag*{[dt16.$i\hspace{.4mm}]_{i=0}^8$}
S(-(x\concd i'))&=-(x\concd i)
\\[2mm]
\label{dt17}
\tag*{[dt17]}
1\concd(-1) &= 9
\\
\label{dt18}
\tag*{[dt18.$i\hspace{.4mm}]_{i=1}^8$}
1\concd(-i') &= i^*
\\
\label{dt19}
\tag*{[dt19.$i\hspace{.4mm}]_{i=1}^8$}
i'\concd(-1) &= i\concd 9
\\
\label{dt20}
\tag*{[dt20.$i.j\hspace{.4mm}]_{i,j=1}^8$}
i'\concd(-j') &= i\concd j^*
\\
\label{dt21}
\tag*{[dt21.$i\hspace{.4mm}]_{i=1}^9$}
(x\concd 0)\concd(-i) &= (x\concd (-1))\concd i^*
\hspace{-22mm}
\\
\label{dt22}
\tag*{[dt22.$i.j\hspace{.4mm}]_{i,j=0}^8$}
\\
\nonumber
(x\concd i')\concd(-j') &= (x\concd i)\concd (j')^*
\\[2mm]
\label{dt23}
\tag*{[dt23]}
(-x)\concd  y&=-(x\concd (-y))
\\
\label{dt24}
\tag*{[dt24.$i\hspace{.4mm}]_{i=0}^9$}
x\concd  (-(y\concd  i))&=-(((-x)+y)\concd  i)
\hspace{-22mm}
\\[3mm]
\label{dt25}
\tag*{[dt25]}
x+(-y)&=-((-x)+y)
\hspace{-22mm}
\\[2mm]
\label{dt26}
\tag*{[dt26]}
x\cdot (-y)&=-(x\cdot y)
\end{Lalign}
\end{minipage}
\hrule
\caption{DDRSs $\nname{dt}$ for~$\Nat_{dt}$ (left column) and
$\zname{dt}$ for~$\Int_{dt}$, for natural number and integer arithmetic with decimal digit 
tree constructor (using the notations $i^\prime$ from Table~\ref{fig:enumdig} and
$i^\star$ from Table~\ref{fig:subdig})} 
\label{tab:intdecview}
\end{table}

Observe that 
equations~\ref{dt2} are instances of their binary counterpart~\ref{bt2}
(see Table~\ref{tab:intbview}), although the AProVE tool~\cite{AProVE} finds that  
the DDRS $\nname{dt}$ is terminating if we replace~\ref{dt2} by
\[x\concd(y\concd z)=(x+ y)\concd z,\]
and also a ground-confluence proof can be easily given. The reason for this 
replacement concerns the generalization to integer arithmetic, as discussed below.

\medskip
The extension to integers is given by the DDRS $\zname{dt}$ in 
Table~\ref{tab:intdecview}, which defines the datatype $\Int_{dt}$.
Clearly, two distinct closed normal forms have distinct values in $\Int$.
Furthermore, the equations in Table~\ref{tab:intdecview} are semantic 
consequences of the axioms for commutative rings  (Table~\ref{tab:CR}).

In contrast to the approaches in \textsc{Walters}~\cite{Wal94} and 
\textsc{Walters \& Zantema}~\cite{WZ95}
with juxtaposition, we now make use of successor terms, and the DDRS presented here 
is composed from rewrite rules for successor, rewrite rules defined in~\cite{Wal94} 
and~\cite{WZ95}, and combinations thereof. 

The above-mentioned equation $x\concd(y\concd z)=(x+ y)\concd z$ is replaced by
the ten 
equations~\ref{dt2} because we failed to find a TRS that employed this equation
and that could be proven terminating by the AProVE tool.
For uniformity, we also replaced the decimal counterpart of equation~\ref{bt18}
(see Table~\ref{tab:intbview}), 
that is
\[x\concd (-(y\concd z))=-((y+(-x))\concd z)\]
by the ten equations~\ref{dt24}. 

\begin{thm}
\label{thm:5.3}
The DDRSs $\nname{dt}$ and $\zname{dt}$ (Table~\ref{tab:intdecview}) are 
ground-complete.
\end{thm}

\begin{proof}
The AProVE tool~\cite{AProVE} finds that both these DDRSs are terminating.
In Appendix~\ref{app:B.2} we prove that the DDRS $\zname{dt}$ is ground-confluent. 

Ground-confluence of $\nname{dt}$ follows in a similar way by restricting the 
proof to the set of nonnegative normal forms.
\end{proof}

We finally note that when we convert the DDRSs from Table~\ref{tab:intdecview} 
to base 2, we obtain alternative DDRSs for the canonical term algebras
$\Nat_{bt}$ and $\Int_{bt}$ that are also ground-complete.

\section{Conclusions}
\label{sec:6}
This paper is about defining (by means of trial and error)  
DDRSs for natural number and integer arithmetic rather than about the precise 
analysis of the various rewriting systems per se.  
What matters in addition to readability and conciseness of each DDRS is at this stage 
a proof that it is terminating and ground-confluent (and thus ground-complete), 
and furthermore that the (intended) normal forms are natural and convincing, while the rewriting
systems are comprehensible.

\medskip
In Section~\ref{sec:2} we provided two DDRSs for the datatype $\Int_r$,
the ring of integers with the set $\NF$ of normal forms defined by
\[
\NF=\{0\}\cup\PNF\cup\NNF,\quad
\PNF= \{1\}\cup\{t+1\mid t\in\PNF\},\quad
\NNF=\{-t\mid t\in\PNF\}.
\]
Each of these DDRSs consists of twelve equations. Perhaps the DDRS $\dname2$
is most attractive: it is comprehensible and deterministic 
with respect to addition of nonnegative closed normal forms.
We leave it as an open question whether $\Int_r$ can be specified 
by a DDRS with fewer equations (preserving the set $\NF$ as normal forms).
Another open question is to find a DDRS for $\Int_r$ and $\NF$
that is also deterministic with respect to rewriting $t\cdot t'$ for nonnegative closed
normal forms $t$ and $t'$.\footnote{Note that the alternative for equation~\ref{R6} 
  suggested by the DDRS $\zname2$, 
  does not solve this open question.
  }
One more open question is whether $\Int_r$ can be specified by a complete term 
rewriting system with the same normal forms.

\medskip
In Section~\ref{sec:3} we provided two DDRSs for natural number and integer arithmetic in 
unary view, based on the constant 0 and unary append. 
Both DDRSs for integer arithmetic contain only nine
equations and we leave it as an open question whether the resulting canonical term algebra 
$\Int_U$ can be specified as a DDRS with fewer equations. 
Concerning their counterparts that define natural number arithmetic ($\nname1$ and
$\nname2$, both containing four equations), the DDRS $\nname1$ is deterministic with respect 
to addition and multiplication of closed normal forms 
(Proposition~\ref{prop:3.2}). Furthermore, the DDRSs $\nname1$ and $\nname2$
are attractive, if only from a didactical point of view:
\begin{enumerate}
\item
Positive numbers are directly related to tallying and admit 
an easy representation and simplifying abbreviations for normal forms, 
such as 011 for $(0\apue)\apue$,
or even 11 or $||$ when removal of the leading zero in positive numbers is adopted. 
\item 
Natural number arithmetic on small numbers can be represented in a comprehensible way
that is fully independent of the learning of any positional system for number 
representation, although  names of numbers (zero, one, two, and so on) might be very 
helpful.\footnote{In English, Dutch and German, this naming is up to twelve
  independent of decimal representation, and in French this is up to sixteen.}
Furthermore, notational abbreviations for units of five, like in
\\[2mm]
\begin{picture}(0,1)(-10,0)
\put(10,0){1111}
\put(6,0){\line(4,1){26}}
\put(40,0){1111}
\put(36,0){\line(4,1){26}}
\put(70,0){11}
\put(90,0){or}
\put(120,0){$||||$}
\put(117,-2){\line(2,1){18}}
\put(140,0){$||||$}
\put(137,-2){\line(2,1){18}}
\put(160,0){$||$}
\put(176,0){or}
\put(196,0){$011111~11111~11$}
\end{picture}
\\[2mm]
can be helpful because 
$0111111111111$ (thus twelve)
is not very well readable or easily distinguishable from
$011111111111$ (thus eleven).
\end{enumerate}
With respect to negative numbers, similar remarks can be made, 
but 
displaying computations according to the DDRSs $\zname1$ and $\zname2$
will be more complex and bracketing seems to be unavoidable.
Consider for example
\begin{align*}
\begin{picture}(110,1)(90,0)
\put(96,0){$(-(\;||\;))~+$}
\put(150,0){$||||$}
\put(147,-2){\line(2,1){18}}
\put(170,0){$||||$}
\put(167,-2){\line(2,1){18}}
\put(190,0){$||$}
\end{picture}
&=
\begin{picture}(120,1)(90,0)
\put(96,0){$(-(\;||\;)\;|\;)~+$}
\put(160,0){$||||$}
\put(157,-2){\line(2,1){18}}
\put(180,0){$||||$}
\put(177,-2){\line(2,1){18}}
\put(200,0){$|$}
\end{picture}
=
\begin{picture}(110,1)(90,0)
\put(96,0){$(-(\;|\;))~+$}
\put(150,0){$||||$}
\put(147,-2){\line(2,1){18}}
\put(170,0){$||||$}
\put(167,-2){\line(2,1){18}}
\put(190,0){$|$}
\end{picture}
=
~\dots
\end{align*}

Although it can be maintained that as a constructor, unary append
is a more illustrative notation than the successor function, it is of course
only syntactic sugar for that function. 

Furthermore, if we add the predecessor function $P(x)$ to the DDRSs defined in 
Section~\ref{sec:3} by the three equations
\[\text{[P1]\hspace{3mm}$P(0)=-(0\apue)$,\quad 
[P2]\hspace{3mm}$P(x\apue) = x$,\quad
[P3]\hspace{3mm}$P(-x) = -(x\apue)$
\hspace{3.32cm}}
\]
we find that the resulting DDRSs improve on those for unary view 
defined in~\cite{BP16a} in terms of simplicity and number of equations.
Finally, if we also add the subtraction function $x - y$ by the single equation
\[\text{[Sub]\hspace{3mm}$x-y= x+(-y)$
\hspace{11.06cm}}
\]
this also improves on the specification of integer arithmetic in unary view 
in \textsc{Walters \& Zantema}~\cite{WZ95}, which does not employ the unary minus function and 
contains seventeen rules, 
and thus eighteen rules when adding the minus function by the rewrite rule 
\[-x~\to~0-x.\]
However, we should mention that the normal forms for negative numbers
in~\cite{WZ95} are $P(0), P(P(0)),...$,
instead of $-(0\apue), -((0\apue)\apue),...$. 

\medskip
In Section~\ref{sec:4} we considered the question how to specify a datatype of integers as an extension 
of the naturals as specified in~\cite{Bergstra2014n}. In this case, the unary view
leads to satisfactory results, but with high inefficiency. 
For the binary view and the decimal view based on the unary append functions 
as discussed in this section, such 
extensions are provided, but the resulting rewriting systems are at first sight 
significantly less concise and comprehensible. 
Some additional notes:

\begin{enumerate}
\item The three DDRSs for integers given in 
Section~\ref{sec:4} each produce an extension datatype for a datatype for the
natural numbers. 
An initial algebra specification of the datatype of integers is obtained from 
any of the DDRSs given in~\cite{Bergstra2014n} by 
\begin{itemize}
\item taking the reduct to the signature involving unary, binary, and decimal notation only,
\item expanding the signature with a unary additive inverse and a unary predecessor function,
\item adding rewrite rules (in equational form) that allow for the unique normalisation of 
closed terms involving the minus sign,
while making sure that these rewrite rules (viewed as equations) are semantic 
consequences of the equations for commutative rings.
\end{itemize}
\item The DDRSs for the binary view and the decimal view 
are hardly intelligible unless one knows that the objective is to construct a 
commutative ring. 
A decimal normal form is defined as either a 
digit, or an application of a decimal append function $\_\apd i$ to a nonzero normal form 
(for all digits $i$). 
This implies the absence of (superfluous) leading zeros, and the closed normal forms thus 
obtained correspond bijectively to the nonnegative integers (that is, $\Nat$).
Incorporating all minus instances $-(t)$ of each nonzero normal form $t$ yields
the class of closed normal forms.

\item Understanding the concept of a commutative ring can be expected only from a person who has
already acquired an understanding of the structure of integers and who accepts the concept of 
generalization of a structure to a class of structures sharing some but not all of its properties.

In other words, the understanding that a DDRS for the integers is provided in the binary view 
and in the decimal view can only
be communicated to an audience under the assumption that a reliable mental picture 
of the integers already exists in the minds of members of the audience. 
This mental picture, however, can 
in principle be communicated by taking notice of the DDRS for the unary view first.
This conceptual (near) circularity may nevertheless be considered a significant 
weakness of the approach of defining (and even introducing) the
integers as an extension of naturals by means of rewriting.
\end{enumerate}

\medskip

In Section~\ref{sec:5} we discussed some alternatives for the above-mentioned DDRSs based on
papers of \textsc{Bouma \& Walters}~\cite{BW89},
\textsc{Walters}~\cite{Wal94}, and \textsc{Walters} and \textsc{Zantema}~\cite{WZ95} 
in which digit tree constructors are used.
In~\cite{Wal94}, \textsc{Walters} presents a TRS 
based on juxtaposition as a tree constructor for integer arithmetic with addition and subtraction 
that is ground-complete and parametric over any radix.
In~\cite{WZ95}, \textsc{Walters} and \textsc{Zantema} extend this TRS with multiplication 
and prove ground-completeness, using 
\emph{semantic labelling} for their termination proof, and judge this TRS | named JP
(juxtaposition) | to have good efficiency 
and readability (in comparison with some alternatives discussed in that paper). 
In~\cite{KW16}, \textsc{Kluiving} and \textsc{van Woerkom}
showed termination of the two DDRSs $\nname{bt}$ and $\zname{bt}$ for arithmetic over $\Nat$ and 
$\Int$ that employ 
the binary tree constructors (Table~\ref{tab:intbview}) with the tool \text{AProVE}~\cite{AProVE}. 
Furthermore, they
also proposed a TRS for arithmetic over the natural numbers employing decimal
tree constructors and proved termination with~\text{AProVE}. 
However, its natural extension to a TRS for integer arithmetic could not be proven terminating, 
probably due to its size.
This finally led us to the DDRSs $\nname{dt}$ and $\zname{dt}$ (Table~\ref{tab:intdecview}).
We leave it as an open question whether $\Int_{dt}$ (the datatype defined by $\zname{dt}$)
can be specified with fewer equations in such a way that a termination proof can be found with 
AProVE (or another tool).

\medskip

A general property of the DDRSs defined in this paper is that the recursion in the 
definitions of addition and multiplication 
takes place on the right argument of these operators (as is common),
if necessary first replacing negation. 
We could have used recursion on the left 
argument instead, obtaining symmetric versions of these DDRSs (for natural number
arithmetic with successor function, this is done in e.g.~\textsc{Bouma} \& 
\textsc{Walters}~\cite{BW89}, \textsc{Walters}~\cite{Walters90},
and \textsc{Zantema}~\cite{Zantema03}).

Of course, 
many normal forms in decimal notation have \emph{names} that confirm their base, 
for example
``six hundred eighty-nine'' $\langle\textsc{ae}\rangle$ or 
``six hundred and eighty-nine'' $\langle\textsc{be}\rangle$.
A decimal notation as 689 is so common that one usually does not question whether it 
represents $(6\apd 8)\apd 9$ or $(6\concd 8)\concd 9$ or some other formally defined notation.
Nevertheless, as we have seen, different algorithmic approaches to for example addition
may apply, although one would preferably not hamper an (initial) arithmetical method with notation 
such as $x\concd (y\concd z)$ and rewrite rules such as $x\concd (y\concd z)\to(x+ y)\concd z$, 
and for this reason we have a preference for
the DDRSs that employ the various append constructors. 

\begin{table}
~\\[0mm]
\centering
\begin{tabular}{|l||r|r|r||r|r|r|r|}
\hline
&\phantom{\hspace{16mm}}&\phantom{\hspace{16mm}}&\phantom{\hspace{16mm}}
&\phantom{\hspace{16mm}}&\phantom{\hspace{16mm}}&\phantom{\hspace{16mm}}
\\[-2mm]
Name
&
$\nname{{dub}}$
&
$\nname{{dt}}$
&DA(10)
&
$\zname{dub}$
&
$\zname{dt}$
&
JP(10)
\\[2mm]
\hline
&&&&&&\\[-2mm]
rule count
&70
&71
&235
&124
&276
&438
\\[2mm]
rule schemes
&12
&11
&10
&26
&26
&30
\\[2mm]
extra operators
&$S(x)$
&$S(x)$
&$s(x), \star_{\delta^0}$
&$S(x)$, $P(x)$
&$S(x)$
&$x-y$
\\[2mm]
\hline
&&&&&&\\[-2mm]
{\small Table / reference}&
{\small \ref{tab:indecview}}
&
{\small \ref{tab:intdecview}}
&
{\small\cite{WZ95}}
&
{\small \ref{tab:indecview}}
&
{\small \ref{tab:intdecview}}
&
{\small\cite{WZ95}}
\\[1mm]
\hline
\end{tabular}
\caption{Rule count for decimal natural and integer arithmetic, where 
DA(10) and JP(10) both originate from~\cite{WZ95} instantiated for base 10.} 
\label{tab:summa}
\end{table}

In Table~\ref{tab:summa} we present the rule count of the DDRSs for 
decimal representation of natural number and integer arithmetic defined in this paper,
and that of the TRSs considered in \textsc{Walters} \& \textsc{Zantema}~\cite{WZ95} 
(the TRS named DA is based on digit append constructors and is discussed below). 
Note that  we do \emph{not} count equations that define the informal
extra operators such as $\boldsymbol{+}^{\!10}$ (Table~\ref{tab:indecview}) and 
$\sumi^1, ...,\sumi^9$ (Table~\ref{tab:intdecview}), which were used only to get a readable table. 
For a fair comparison in the case of the decimal append functions we also leave out the 
equations for conversion to binary view. 
This shows that our DDRSs are relatively concise, but we note that we did not succeed in getting
a more extended overview on specifications for natural number and integer arithmetic, and 
that it might well be that we missed relevant TRSs.

For each DDRS it may be taken for a quality criterion if normalising reductions are not excessively long. 
We leave it as an open question for each of the systems presented to find meaningful 
upper bounds (in terms of $n$) of the number of steps of the longest reduction which can be made from a 
ground term of size $n$. Especially for the cases of binary and decimal numbers it is interesting to 
compare these values with the number of steps needed for instance when performing leftmost innermost 
normalisation where addition and multiplication are performed by means of the standard (``school") 
algorithms, or even with faster algorithms like Sch\"onhage-Strassen for multiplication.

We have proven that all DDRSs considered in this paper are ground-complete, and 
both their termination
proofs and all associated TRSs used for these proofs can be found at
\\
\url{https://arxiv.org/src/1608.06212/anc/}.
It should be noted that handwritten ground-confluence proofs of the size recorded in 
Appendices~\ref{app:A} and~\ref{app:B} are error-prone; however, our proofs can be automated 
as was shown by \textsc{van Woerkom} in~\cite{Woerkom17}, which also contains a general theorem
about this type of ground-confluence proofs (Thm.1).
Furthermore, with the confluence prover CSI~\cite{csi}, 
confluence proofs were found for the DDRSs for natural number arithmetic 
with unary append $\nname1$ (Table~\ref{tab:intuview1})
and $\nname 2$ (Table~\ref{tab:int2}), thus these DDRSs are complete.
For all other DDRSs defined in this paper, confluence can be disproven by CSI. 
Finally, the tool AGCP~\cite{ATK} found ground-confluence proofs
for the DDRSs for natural number arithmetic with unary append 
$\nname{ubd}$ (Table~\ref{tab:natunview})
and with unary digit tree constructor $\nname{ut}$ (Table~\ref{tab:intuaview}),
and also for the DDRSs $\nname1$ and $\nname 2$, but for none of the remaining DDRSs.

\medskip

We briefly discuss two other, comparable approaches to arithmetic that are also based on
some form of digit append constructors for representing  numbers. 
First, in~\cite{WZ95} \textsc{Walters} and \textsc{Zantema} introduce a TRS which they
named DA (for ``digit application'')
with addition and multiplication on natural numbers. The authors prove termination by 
recursive path ordering and confluence, and also judge this TRS to have 
good efficiency and readability. 
Secondly, in~\cite{CMR97}, \textsc{Contejean, March\'e} and \textsc{Rabehasaina} 
introduce integer arithmetic based on
\emph{balanced ternary numbers}, that is, numbers that can be represented by a digit append
function $\apt$ 
with digits $-1,0,1$ and semantics $\llbracket i\rrbracket= i$ and 
$\llbracket x\apt i\rrbracket= 3\cdot\llbracket x\rrbracket+i$ (see, e.g., 
\textsc{Knuth}~\cite{Knuth})
and provide a TRS that is confluent and terminating modulo associativity and commutativity of 
addition and multiplication.

Based on either a DDRS for the natural numbers or a DDRS for the integers one may develop a 
DDRS for rational numbers in various ways.
It is plausible to consider the meadow of rational numbers of~\cite{BergstraT07} or the
non-involutive meadow of rational numbers (see \cite{BM2014}) or the common meadow of rational 
numbers (see \cite{BP2014}) as abstract algebraic structures for rationals in which unary, 
binary, and decimal notation are to be incorporated in ways possibly based on the specifications 
presented above. 
Furthermore, one does well to consider the work discussed in~\cite{CMR97}
on a TRS for rational numbers, in which
arithmetic for rational numbers is specified (this is the main result in~\cite{CMR97}, for which 
the above-mentioned work on integer arithmetic is a preliminary): the authors specify rational 
numbers by means of a TRS that is complete modulo associativity and commutativity of addition 
and multiplication, taking advantage of Stein's algorithm for computing gcd's of nonnegative 
integers without any division\footnote{%
  Apart from halving even numbers, which is easy in binary notation, but
  can otherwise be specified with a shift operation.}
(see, e.g., \cite{Knuth}).

A survey of equational algebraic specifications for abstract datatypes is provided
by \textsc{Wirsing}~\cite{Wirsing91}.
In~\cite{BergstraT95} one finds the general result that computable abstract datatypes 
can be specified by means of
specifications which are confluent and terminating term rewriting systems. 
Some general results on algebraic specifications can be found 
in~\cite{BroyWP84,BergstraT87,GaudelJ98}. 
More recent applications of equational specifications can be 
found in~\cite{BergstraT07}.

\section*{Acknowledgement} 
Many thanks to Wijnand van Woerkom for carefully identifying errors and gaps in an
earlier version of this work, and for some very useful suggestions, including  
those for better rewrite rules for the DDRS $\zname{dt}$ (Table~\ref{tab:intdecview}). 
Furthermore, his help in proving termination 
of some of the larger DDRSs with the AProVE tool and providing these proofs was 
crucial\footnote{%
  For example, we were able to use AProVE's web interface to prove termination of 
  the DDRS $\nname{dt}$, but for the DDRS $\zname{dt}$ this was only 
  possible after reducing this DDRS to base 2.}
and led to the webarchive~\url{https://arxiv.org/src/1608.06212/anc/}.
Also, many thanks to three reviewers for their careful and
comprehensive reports, and very helpful suggestions.

\appendix

\section{DDRSs with digit append constructors, ground-confluence}
\label{app:A}
In this appendix we prove ground-confluence of the DDRSs
$\zname{bud}$ and $\zname{dub}$. 
In both proofs we adopt the approach used in the proof of Theorem~\ref{thm:2.2}.

\subsection{Binary view: $\zname{bud}$, the DDRS for $\Int_{bud}$}
\label{app:A.1}
~\\
This DDRS is defined in Table~\ref{tab:binview}.
Define the set $N$ of closed terms over $\Sigma_\Int$ as follows:
\begin{align*}
N&=\{0\}\cup N^+\cup N^-,\\
N^+&=\{1\}\cup\{t\apb 0,t\apb 1\mid t\in N^+\},\\
N^-&=\{-t\mid t\in N^+\}.
\end{align*}
It immediately follows that if $t\in N$, then $t$ is a normal form (no rewrite 
rule applies).

In order to prove ground-confluence of this rewriting system, it suffices to show that
for each closed term $t$ over $\Sigma_\Int$, either $t\in N$ or $t$ has a rewrite step, 
so that each normal form is in $N$. 
We prove this by structural induction on $t$.

The base cases are simple: if $t\in\{0,1\}$ then  $t\in N$, and 
if $t=i'$ for some $i\in\{1,2,3,4,5,6,7,8\}$, then $t$ has a rewrite step by 
equation~[\textbf{bd1.$i$}].
For the induction step we distinguish eight cases:
\begin{enumerate}

\item 
Case $t=S(r)$. Assume that $r\in N$ and apply case distinction on $r$:
\begin{itemize}
\item
if $r=0$, then $t$ has a rewrite step by equation~\ref{b2},
\item
if $r=1$, then $t$ has a rewrite step by equation~\ref{b3},
\item
if $r=r'\apb 0$, then $t$ has a rewrite step by equation~\ref{b4},
\item
if $r=r'\apb 1$, then $t$ has a rewrite step by equation~\ref{b5},
\item
if $r=-1$, then $t$ has a rewrite step by equation~\ref{b20},
\item
if $r=-(r'\apb 0)$, then $t$ has a rewrite step by equation~\ref{b21},
\item 
if $r=-(r'\apb 1)$, then $t$ has a rewrite step by equation~\ref{b22}.
\end{itemize}
\item
Case $t=P(r)$. Assume that $r\in N$ and apply case distinction on $r$:
\begin{itemize}
\item
if $r=0$, then $t$ has a rewrite step by equation~\ref{b15},
\item
if $r=1$, then $t$ has a rewrite step by equation~\ref{b16},
\item
if $r=r'\apb 0$, then $t$ has a rewrite step by equation~\ref{b17},
\item
if $r=r'\apb 1$, then $t$ has a rewrite step by equation~\ref{b18},
\item
if $r\in N^-$, then $t$ has a rewrite step by equation~\ref{b19}.
\end{itemize}
\item 
Case $t=-r$. Assume that $r\in N$ and apply case distinction on $r$:
\begin{itemize}
\item
if $r=0$, then $t$ has a rewrite step by equation~\ref{b13},
\item
if $r=1$, then $t\in N$,
\item
if $r=r'\apb i$, then $t\in N$,
\item
if $r\in N^-$, then $t$ has a rewrite step by equation~\ref{b14}.
\end{itemize}
\item
Case $t=r\apb 0$. Assume that $r\in N$ and apply case distinction on $r$:
\begin{itemize}
\item
if $r=0$, then $t$ has a rewrite step by equation~[b1.0],
\item
if $r\in N^+$, then $t\in N$,
\item
if $r\in N^-$, then $t$ has a rewrite step by equation~\ref{b23}.
\end{itemize}
\item
Case $t=r\apb 1$. Assume that $r\in N$ and apply case distinction on $r$:
\begin{itemize}
\item
if $r=0$, then $t$ has a rewrite step by equation~[b1.1],
\item
if $r\in N^+$, then $t\in N$,
\item
if $r\in N^-$, then $t$ has a rewrite step by equation~\ref{b24}.
\end{itemize}
\item 
Case $t=u+r$. Assume that $u,r\in N$ and apply case distinction on $r$:
\begin{itemize}
\item
if $r=0$, then $t$ has a rewrite step by equation~\ref{b6},
\item
if $r=1$, then $t$ has a rewrite step by equation~\ref{b7},
\item 
if $r=r'\apb 0$, then $t$ has a rewrite step according to equation~\ref{b8},
\item 
if $r=r'\apb 1$, then $t$ has a rewrite step according to equation~\ref{b9},
\item
if $r\in N^-$, then $t$ has a rewrite step according to equation~\ref{b25}.
\end{itemize}
\item 
Case $t=u\cdot r$. Assume that $u,r\in N$ and apply case distinction on $r$:
\begin{itemize}
\item
if $r=0$, then $t$ has a rewrite step by equation~\ref{b10},
\item
if $r=1$, then $t$ has a rewrite step by equation~\ref{b11},
\item 
if $r=r'\apb i$, then $t$ has a rewrite step according to~[b12.$i$],
\item
if $r\in N^-$, then $t$ has a rewrite step by equation~\ref{b26}.
\end{itemize}
\item
Case $t=r\apd i$ for $i\in D$.
Now $t$ has a rewrite step by equation~[\textbf{bd2.$i$}].
\end{enumerate}
This concludes our proof.

\subsection{Decimal view: $\zname{dub}$, the DDRS for $\Int_{dub}$}
\label{app:A.2}
~\\
This DDRS is defined in Table~\ref{tab:indecview}.
Recall that we write $D$  for the set of all digits.  

Define the set $N$ of closed terms over $\Sigma_\Int$ as follows:
\begin{align*}
N&=\{0\}\cup N^+\cup N^-,\\
N^+&=D\setminus\{0\}\cup\{t\apd i\mid t\in N^+, i\in D\},\\
N^-&=\{-t\mid t\in N^+\}.
\end{align*}
It immediately follows that if $t\in N$, then $t$ is a normal form (no rewrite 
rule applies).
In order to prove ground-confluence of this rewriting system, it suffices to show that
for each closed term $t$ over $\Sigma_\Int$, either $t\in N$ or $t$ has a rewrite step, 
so that each normal form is in $N$. 
We prove this by structural induction on $t$.

The base cases are trivial: if $t\in D$, then  $t\in N$.
For the induction step we distinguish eight cases:
\begin{enumerate}

\item 
Case $t=S(r)$. Assume that $r\in N$ and apply case distinction on $r$:
\begin{itemize}
\item
if $r=i$ for $i\in\{0,1,\dots,8\}$, then $t$ has a rewrite step by equation~[d2.$i$],
\item
if $r=9$, then $t$ has a rewrite step by equation~\ref{d3},
\item
if $r=r'\apd i$ for $i\in\{0,1,\dots,8\}$, then $t$ has a rewrite step by equation~[d4.$i$],
\item
if $r=r'\apd 9$, then $t$ has a rewrite step by equation~\ref{d5},
\item
if $r=-i'$ for $i\in\{0,1,\dots,8\}$, then $t$ has a rewrite step by equation~[d20.$i$],
\item
if $r=-(r'\apd 0)$, then $t$ has a rewrite step by equation~\ref{d21},
\item 
if $r=-(r'\apd i')$ for $i\in\{0,1,\dots,8\}$, then $t$ has a rewrite step 
by equation~[d22.$i$].
\end{itemize}
\item
Case $t=P(r)$. Assume that $r\in N$ and apply case distinction on $r$:
\begin{itemize}
\item
if $r=0$, then $t$ has a rewrite step by equation~\ref{d15},
\item
if $r=i'$ for $i\in\{0,1,\dots,8\}$, then $t$ has a rewrite step by equation~[d16.$i$],
\item
if $r=r'\apd 0$, then $t$ has a rewrite step by equation~\ref{d17},
\item
if $r=r'\apd i'$ for $i\in\{0,1,\dots,8\}$, then $t$ has a rewrite step by equation~[d18.$i$],
\item
if $r\in N^-$, then $t$ has a rewrite step by equation~\ref{d19}.
\end{itemize}
\item 
Case $t=-r$. Assume that $r\in N$ and apply case distinction on $r$:
\begin{itemize}
\item
if $r=0$, then $t$ has a rewrite step by equation~\ref{d13},
\item
if $r\in N^+$, then $t\in N$,
\item
if $r\in N^-$, then $t$ has a rewrite step by equation~\ref{d14}.
\end{itemize}

\item
Case $t=r\apd 0$. Assume that $r\in N$ and apply case distinction on $r$:
\begin{itemize}
\item
if $r=0$, then $t$ has a rewrite step by equation~[d1.0],
\item
if $r\in N^+$, then $t\in N$,
\item
if $r\in N^-$, then $t$ has a rewrite step by equation~\ref{d23}.
\end{itemize}
\item
Case $t=r\apd i$ for $i\in\{1,2,\dots,9\}$. Assume that $r\in N$ and apply case distinction on $r$:
\begin{itemize}
\item
if $r=0$, then $t$ has a rewrite step by equation~[d1.$i$],
\item
if $r\in N^+$, then $t\in N$,
\item
if $r\in N^-$, then $t$ has a rewrite step by equation~[d24.$i$].
\end{itemize}
\item 
Case $t=u+r$. Assume that $u,r\in N$ and apply case distinction on $r$:
\begin{itemize}
\item
if $r=0$, then $t$ has a rewrite step by equation~\ref{d6},
\item
if $r=i$ for $i\in\{1,2,\dots,9\}$, then $t$ has a rewrite step by equation~[d7.$i$],
\item 
if $r=r'\apd 0$, then $t$ has a rewrite step by equation~\ref{d8},
\item 
if $r=r'\apd i$ for $i\in\{1,2,\dots,9\}$, then $t$ has a rewrite step by  
equation~[d9.$i$],
\item
if $r\in N^-$, then $t$ has a rewrite step by equation~\ref{d25}.
\end{itemize}
\item 
Case $t=u\cdot r$. Assume that $u,r\in N$ and apply case distinction on $r$:
\begin{itemize}
\item
if $r=0$, then $t$ has a rewrite step by equation~\ref{d10},
\item
if $r=i'$ for $i\in\{0,1,\dots,8\}$, then $t$ has a rewrite step by~[d11.$i$],
\item
if $r=r'\apd i$ for $i\in D$, then $t$ has a rewrite step by equation~[d12.$i$],
\item 
if $r\in N^-$, then $t$ has a rewrite step by equation~\ref{d26}.
\end{itemize}
\item
Case $t=r\apb i$ for $i\in\{0,1\}$.
Now
$t$ has a rewrite step by equation~[\textbf{db1.$i$}].
\end{enumerate}
This concludes our proof.

\section{DDRSs with digit tree constructors, ground-confluence}
\label{app:B}
In this appendix we prove ground-confluence of the DDRSs $\zname{ut}$, 
$\zname{bt}$, 
and $\zname{dt}$, respectively.
In all proofs we adopt the approach used in the proof of Theorem~\ref{thm:2.2}.

\subsection{Unary view: the DDRS $\zname{ut}$ for $\Int_{ut}$}
\label{app:B.1}
~\\
This DDRS is defined in Table~\ref{tab:intuaview}. 
Define the signature $\Sigma_{ut}=\{0,-(\_),~\_\concu\_~,+,\cdot\}$, and
the set $N$ of closed terms over $\Sigma_{ut}$ as follows:
\begin{align*}
N&=\{0\}\cup N^+\cup N^-,\\
N^+&=\{0\concu 0\}\cup \{t\concu 0\mid t\in N^+\},\\
N^-&=\{-t\mid t\in N^+\}.
\end{align*}
It immediately follows that if $t\in N$, then $t$ is a normal form (no rewrite 
rule applies).
In order to prove ground-confluence, it suffices to show that
for each closed term $t$ over $\Sigma_{ut}$, either $t\in N$ or $t$ has a rewrite step, 
so that each normal form is in $N$. 
We prove this by structural induction on $t$.

The base case is trivial: if $t=0$, then  $t\in N$.
For the induction step we distinguish four cases:
\begin{enumerate}
\item 
Case $t=-r$. Assume that $r\in N$ and apply case distinction on $r$:
\begin{itemize}
\item
if $r=0$, then $t$ has a rewrite step by equation~\ref{ut6},
\item
if $r=r'\concu 0$, then $t\in N$,
\item 
if $r=-(r'\concu 0)$, then $t$ has a rewrite step by equation~\ref{ut7}.
\end{itemize}
\item
Case $t=v\concu r$. Assume that $v,r\in N$ and apply case distinction on $r$:
\begin{itemize}
\item
if $r=0$, then apply case distinction on $v$:
\begin{itemize}
\item
if $v=0$, then $t\in N$,
\item
if $v=v'\concu 0$, then $t\in N$,
\item
if $v=-(v'\concu 0)$, then $t$ has a rewrite step by equation~\ref{ut10}.
\end{itemize}
\item 
if $r=r'\concu 0$, then apply case distinction on $v$:
\begin{itemize}
\item
if $v=0$, then $t$ has a rewrite step by equation~\ref{ut1},
\item
if $v=v'\concu 0$, then $t$ has a rewrite step by equation~\ref{ut1},
\item
if $v=-(v'\concu 0)$, then $t$ has a rewrite step by equation~\ref{ut11}.
\end{itemize}
\item 
if $r=-(r'\concu 0)$, then apply case distinction on $v$:
\begin{itemize}
\item
if $v=0$, then $t$ has a rewrite step by equation~\ref{ut8},
\item
if $v=v'\concu 0$, then $t$ has a rewrite step by equation~\ref{ut9},
\item 
if $v=-(v'\concu 0)$, then $t$ has a rewrite step by equation~\ref{ut12}.
\end{itemize}
\end{itemize}
\item 
Case $t=v+r$. Assume that $v,r\in N$ and apply case distinction on $r$:
\begin{itemize}
\item
if $r=0$, then $t$ has a rewrite step by equation~\ref{ut2},
\item
if $r=r'\concu 0$, then $t$ has a rewrite step by equation~\ref{ut3},
\item
if $r=-(r'\concu 0)$, then $t$ has a rewrite step by equation~\ref{ut13}.
\end{itemize}
\item 
Case $t=v\cdot r$. Assume that $v,r\in N$ and apply case distinction on $r$:
\begin{itemize}
\item
if $r=0$, then $t$ has a rewrite step by equation~\ref{ut4},
\item
if $r=r'\concu 0$, then $t$ has a rewrite step by equation~\ref{ut5},
\item
if $r=-(r'\concu 0)$, then $t$ has a rewrite step by equation~\ref{ut14}.
\end{itemize}
\end{enumerate}
This concludes our proof.

\subsection{Binary view: $\zname{bt}$, the DDRS for $\Int_{bt}$}
\label{app:B.2}
~\\
This DDRS is defined in Table~\ref{tab:intbview}.
Define the signature $\Sigma_{bt}=\{0,1,-(\_),~\_\concb\_~,+,\cdot\}$, and
the set $N$ of closed terms over $\Sigma_{bt}$ as follows:
\begin{align*}
N&=\{0\}\cup N^+\cup N^-,\\
N^+&=\{1\}\cup\{t\concb 0,t\concb 1\mid t\in N^+\},\\
N^-&=\{-t\mid t\in N^+\}.
\end{align*}
It immediately follows that if $t\in N$, then $t$ is a normal form (no rewrite 
rule applies), and that two distinct elements in $N$ have distinct values in $\Int$.
In order to prove ground-confluence of this rewriting system, it suffices to show that
for each closed term $t$ over $\Sigma_{bt}$, either $t\in N$ or $t$ has a rewrite step, 
so that each normal form is in $N$. 
We prove this by structural induction on $t$.

The base cases are simple: if $t\in\{0,1\}$, then  $t\in N$.
For the induction step we distinguish four cases:
\begin{enumerate}
\item 
Case $t=-r$. Assume that $r\in N$ and apply case distinction on $r$:
\begin{itemize}
\item
if $r=0$, then $t$ has a rewrite step by equation~\ref{bt13},
\item
if $r\in N^+$, then $t\in N$,
\item
if $r\in N^-$, then $t$ has a rewrite step by equation~\ref{bt14}.
\end{itemize}
\item
Case $t=r\concb  u$. Assume that $r,u\in N$ and apply case distinction on $u$:
\begin{itemize}
\item
if $u\in \{0,1\}$, then apply case distinction on $r$:
\begin{itemize}
\item
if $r=0$, then $t$ has a rewrite step by equation~\ref{bt1},
\item
if $r\in N^+$, then $t\in N$,
\item
if $r\in N^-$, then $t$ has a rewrite step by equation~\ref{bt19}.
\end{itemize}
\item
if $u=u'\concb  j$, then $t$ has a rewrite step by equation~\ref{bt2},
\item 
if $u=-1$, then apply case distinction on $r$:
\begin{itemize}
\item
if $r=0$, then $t$ has a rewrite step by equation~\ref{bt1},
\item
if $r=1$, then $t$ has a rewrite step by~\ref{bt15},
\item 
if $r=r'\concb  0$, then $t$ has a rewrite step by equation~\ref{bt16},
\item 
if $r=r'\concb  1$, then $t$ has a rewrite step by equation~\ref{bt17},
\item
if $r\in N^-$, then $t$ has a rewrite step by equation~\ref{bt19}.
\end{itemize}
\item
if $u=-(u'\concb  j)$, then $t$ has a rewrite step by equation~\ref{bt18}.
\end{itemize}

\item 
Case $t=u+r$. Assume that $u,r\in N$ and apply case distinction on $r$:
\begin{itemize}
\item
if $r=0$, then $t$ has a rewrite step by equation~\ref{bi4},
\item
if $r=1$, then apply case distinction on $u$:
\begin{itemize}
\item
if $u=0$, then $t$ has a rewrite step by equation~\ref{bi3},
\item
if $u=1$, then $t$ has a rewrite step by equation~\ref{bi5}
\item
if $u=u'\concb j$, then $t$ has a rewrite step by equation~\ref{bi7},
\item
if $u=-1$, then $t$ has a rewrite step by equation~\ref{bt21},
\item
if $u=-(u'\concb j)$, then $t$ has a rewrite step by equation~\ref{bt24},
\end{itemize}
\item 
if $r=r'\concb i$, then $t$ has a rewrite step by \ref{bi6},
\item
if $r=-1$, then apply case distinction on $u$:
\begin{itemize}
\item
if $u=0$, then $t$ has a rewrite step by equation~\ref{bi3},
\item
if $u=1$, then $t$ has a rewrite step by equation~\ref{bt20}
\item
if $u=u'\concb j$ then $t$ has a rewrite step by equation~\ref{bi7},
\item
if $u=-1$, then $t$ has a rewrite step by equation~\ref{bt22},
\item
if $u=-(u'\concb j)$, then $t$ has a rewrite step by equation~\ref{bt24}, 
\end{itemize}
\item 
if $r=-(r'\concb i)$, then $t$ has a rewrite step by equation~\ref{bt23}. 
\end{itemize}

\item 
Case $t=u\cdot r$. Assume that $u,r\in N$ and apply case distinction on $r$:
\begin{itemize}
\item
if $r=0$, then $t$ has a rewrite step by equation~\ref{bi8},
\item
if $r=1$, then apply case distinction on $u$:
\begin{itemize}
\item
if $u=0$, then $t$ has a rewrite step by equation\ref{bi9},
\item
if $u=1$, then $t$ has a rewrite step by equation\ref{bt10},
\item
if $u=u'\concb j$, then $t$ has a rewrite step by equation~\ref{bt12},
\item
if $u\in N^-$, then $t$ has a rewrite step by equation~\ref{bt26},
\end{itemize}
\item 
if $r=r'\concb i$, then $t$ has a rewrite step by equation~\ref{bt11},
\item
if $r\in N^-$, then $t$ has a rewrite step by equation~\ref{bt25}.
\end{itemize}
\end{enumerate}
This concludes our proof.

\subsection{Decimal view: $\zname{dt}$, the DDRS for $\Int_{dt}$}
\label{app:B.3}
~\\
This DDRS is defined in Table~\ref{tab:intdecview}.
Recall that $D=\{0,1,2,...,9\}$ and define the signature  
$\Sigma_{dt}=\{+,\cdot,~\_\conct \_~,-(\_),i\mid i\in D\}$, and
the set $N$ of closed terms over $\Sigma_{dt}$ as follows:
\begin{align*}
N&=\{0\}\cup N^+\cup N^-,\\
N^+&=D\setminus\{0\}\cup\{t\conct i\mid t\in N^+, i\in D\},\\
N^-&=\{-t\mid t\in N^+\}.
\end{align*}
It immediately follows that if $t\in N$, then $t$ is a normal form (no rewrite 
rule applies).
In order to prove ground-confluence of this rewriting system, it suffices to show that
for each closed term $t$ over $\Sigma_{dt}$, either $t\in N$ or $t$ has a rewrite step, 
so that each normal form is in $N$. 
We prove this by structural induction on $t$.

The base cases are simple: if $t\in D$, then  $t\in N$.
For the induction step we have to distinguish five cases:
\begin{enumerate}
\item 
Case $t=-r$. Assume that $r\in N$ and apply case distinction on $r$:
\begin{itemize}
\item
if $r=0$ then $t$ has a rewrite step by equation~\ref{dt12},
\item
if $r\in N^+$, then $t\in N$,
\item
if $r\in N^-$, then $t$ has a rewrite step by equation~\ref{dt13}.
\end{itemize}

\item 
Case $t=S(r)$. Assume that $r\in N$ and apply case distinction on $r$:
\begin{itemize}
\item
if $r=i$ for $i\in D\setminus\{9\}$, then $t$ has a rewrite step by 
equation~[dt3.$i$],
\item
if $r=9$, then $t$ has a rewrite step by equation~\ref{dt4},
\item
if $r=r'\conct i$ for $i\in D\setminus\{9\}$, then $t$ has a rewrite step by
equation~[dt5.$i$],
\item
if $r=r'\conct 9$, then $t$ has a rewrite step by equation~\ref{dt6},
\item
if $r=-i'$ for $i\in D\setminus\{9\}$, then $t$ has a rewrite step by 
equation~[dt14.$i$],
\item
if $r=-(r\concd 0)$, then $t$ has a rewrite step by equation~\ref{dt15},
\item
if $r=-(r\concd i')$ for $i\in D\setminus\{9\}$, then $t$ has a rewrite step 
by equation~[dt16.$i$].
\end{itemize}
\item
Case $t=r\conct  u$. Assume that $r,u\in N$ and apply case distinction on $u$:
\begin{itemize}
\item
if $u\in D$, then apply case distinction on $r$:
\begin{itemize}
\item
if $r=0$, then $t$ has a rewrite step by equation~\ref{dt1},
\item
if $r\in N^+$, then $t\in N$,
\item
if $r\in N^-$, then $t$ has a rewrite step by equation~\ref{dt23}.
\end{itemize}
\item
if $u=u'\conct  i$ for $i\in D$, then $t$ has a rewrite step by equation~[dt2.$i$],
\item 
if $u=-i$ for $i\in D\setminus\{0\}$, then apply case distinction on $r$:
\begin{itemize}
\item
if $r=0$, then $t$ has a rewrite step by equation~\ref{dt1},
\item
if $r=1$ and $u=-1$, then $t$ has a rewrite step by equation~\ref{dt17},
\item
if $r=1$ and $u=-i'$ for $i\in D\setminus\{9\}$, then $t$ has a rewrite step by 
equation~[dt18.$i$],
\item
if $r=i'$ and $u=-1$ for $i\in D\setminus\{9\}$, then $t$ has a rewrite step by 
equation~[dt19.$i$],
\item
if $r=i'$ and $u=-j'$ for $i,j\in D\setminus\{9\}$, then $t$ has a rewrite step by 
equation~[dt20.$i.j$],
\item 
if $r=r'\conct 0$ and $u=-i$ for $i\in D\setminus\{0\}$, then $t$ has a rewrite step 
by equation~[dt21.$i$],
\item 
if $r=r'\conct i'$ and $u=-j'$ for $i,j\in D\setminus\{9\}$, then $t$ has a rewrite 
step by equation~[dt22.$i.j$],
\item
if $r\in N^-$, then $t$ has a rewrite step by equation~\ref{dt23}.
\end{itemize}
\item
if $u=-(u'\conct  i)$, then $t$ has a rewrite step by equation~[dt24.$i$].
\end{itemize}
\item 
Case $t=r+u$. Assume that $r,u\in N$ and apply case distinction on $u$:
\begin{itemize}
\item
if $u=i$ for $i\in D$, then $t$ has a rewrite step by equation~[dt7.$i$],
\item 
if $u=u'\conct  i$ for $i\in D$, then $t$ has a rewrite step by equation~[dt8.$i$],
\item
if $u\in N^-$, say $u=-u'$, then $t$ has a rewrite step by equation~\ref{dt25}.
\end{itemize}

\item
Case $t=r\cdot u$. Assume that $r,u\in N$ and apply case distinction on $u$:
\begin{itemize}
\item
if $u=0$, then $t$ has a rewrite step by equation~\ref{dt9},
\item
if $u=i$ for $i\in D\setminus\{0\}$, then $t$ has a rewrite step by equation~[dt10.$i$],
\item 
if $u=u'\concd i$ for $i\in D$, then $t$ has a rewrite step by equation~[dt11.$i$],
\item
if $u\in N^-$, then $t$ has a rewrite step by equation~\ref{dt26}.
\end{itemize}

\end{enumerate}
This concludes our proof.

\end{document}